\documentclass[preprint,10pt]{elsarticle}

\usepackage[a4paper,margin=1in]{geometry}
\usepackage{placeins}
\usepackage{booktabs}

\usepackage{amssymb}
\usepackage{hyperref}
\usepackage{amsmath}
\usepackage{amsthm}
\usepackage{bbm}

\usepackage{algorithm}
\usepackage{algpseudocode}

\usepackage{ragged2e}
\usepackage{array}

\usepackage{caption} 
\usepackage{threeparttable}

\newtheoremstyle{myremark}
  {3pt}
  {3pt}
  {}
  {}
  {\bfseries}
  {.}
  { }
  {}

\theoremstyle{myremark}

\DeclareMathOperator*{\argmax}{arg\,max}

\journal{}

\begin{document}

\begin{frontmatter}



\title{Process–fracture mapping of a DLP-printed photopolymer using Bayesian active learning and surrogate-based sensitivity analysis}

\author[a]{Ethan Blackwell}
\author[b]{Yogesh C. Chandrashekar}
\author[a]{Guoqiang Li}
\author[b]{Kshitiz Upadhyay\corref{cor1}}

\affiliation[a]{organization={Department of Mechanical and Industrial Engineering, Louisiana State University}, addressline={Baton Rouge, LA 70803}, country={USA}}

\affiliation[b]{organization={Department of Aerospace Engineering and Mechanics, University of Minnesota}, addressline={Minneapolis, MN 55455}, country={USA}}

\cortext[cor1]{Corresponding author. Email: kshitizu@umn.edu}

\begin{abstract}
Digital light processing (DLP) enables rapid fabrication of polymer structures, but fracture performance can depend on multiple interacting processing variables, making exhaustive experimental characterization impractical. This work presents a data-efficient framework for process–fracture mapping of a DLP-printed photopolymer using Bayesian active learning and digital image correlation (DIC)–assisted Mode I fracture experiments. Four processing parameters were considered: layer angle, UV exposure time, layer height, and print temperature. Fracture resistance was quantified by the critical J-integral, $J_c$, obtained from three-point-bending tests with DIC-based evaluation of crack-mouth opening displacement and hinge-point kinematics. Beginning with two randomly selected conditions, Gaussian process regression (GPR) and its resulting modified upper confidence bound (UCB)–style acquisition function sequentially selected 26 additional experiments, yielding 28 processing conditions with three replicate specimens per condition. The final GPR surrogate reproduced the training data with $R^2=0.99$ and achieved leave-one-out cross-validation performance of $R^2=0.63$ and Pearson $r=0.81$. Surrogate-based sensitivity analysis was then applied to quantify local parameter effects and global contributions to fracture resistance. One-at-a-time response curves revealed distinctly nonlinear conditional trends, while global Sobol analysis identified UV exposure time as the dominant processing variable, with first-order and total-order indices of 0.6780 and 0.7581, respectively. Based on total-order influence, the parameters ranked as UV exposure time, layer angle, print temperature, and layer height. The first-order Sobol indices summed to 0.8058, indicating non-negligible interaction and higher-order effects. These results demonstrate that Bayesian-active-learning-guided experimentation can efficiently recover informative process–fracture relationships and parameter interactions from a sparse experimental campaign.
\end{abstract}


\end{frontmatter}


\section{Introduction}
\label{sec:Introduction}

Digital light processing (DLP) is a vat-photopolymerization-based additive manufacturing technique in which a projected light pattern selectively cures a liquid photosensitive resin, typically in a layer-by-layer manner, to fabricate three-dimensional parts directly from a digital model \cite{Ngo2018,Pagac2021,AlRashid2021,Chaudhary2023}. Compared with many other polymer additive manufacturing routes, DLP offers an attractive combination of high geometric fidelity, fine spatial resolution, smooth surface finish, and relatively rapid fabrication, since an entire layer can be exposed simultaneously rather than traced point-by-point \cite{Pagac2021,AlRashid2021,Chaudhary2023}. More broadly, additive manufacturing enables the autonomous fabrication of geometrically complex structures, including lattices, graded architectures, and miniaturized devices that are difficult, costly, or in some cases impossible to realize using conventional manufacturing routes. Owing to these advantages, DLP-printed photopolymers have shown growing promise in applications such as soft robotic components, soft sensors and actuators, biomedical devices, and photonic or optoelectronic structures, where design freedom and fine feature resolution are often essential \cite{Ge2022,Gul2018,Zhao2021,Han2019,Chang2025,Chekkaramkodi2024}.

As with most additive manufacturing methods, however, the mechanical performance of DLP-printed polymers depends strongly on processing conditions \cite{Farkas2023,Jiang2023,Lee2018,Brighenti2023}. This sensitivity is expected because DLP is governed by coupled photochemical and manufacturing phenomena: exposure conditions influence polymerization kinetics and degree of cure; curing history affects crosslink density and local network structure; and the layerwise fabrication process can introduce anisotropy, interlayer heterogeneity, and defect sensitivity \cite{Chaudhary2023}. Accordingly, processing variables such as layer thickness, exposure time, print orientation, post-curing conditions, and related thermal or environmental variables can significantly alter stiffness, strength, ductility, dimensional accuracy, and failure behavior of the final part. Representative studies have shown, for example, that tensile, compressive, and flexural properties of DLP-printed polymers vary appreciably with layer thickness and build orientation \cite{Farkas2023,Jiang2023,Lee2018}, while curing-related variables such as UV exposure history can strongly affect both dimensional accuracy and mechanical response through their influence on the extent of polymerization and associated network development \cite{Brighenti2023}.

Among the mechanical properties affected by these parameters, fracture resistance is especially important for DLP-printed photopolymers. Many photopolymer systems used in vat photopolymerization exhibit relatively brittle or glassy behavior under ambient conditions, such that structural performance may be governed not only by nominal strength but also by resistance to crack initiation and propagation in the presence of notches, flaws, or layer-induced heterogeneities \cite{Lenti2019,Khosravani2023}. More generally, recent reviews of additively manufactured components have identified fracture as a central design concern because AM parts often contain process-induced microstructural features, interfacial weaknesses, and geometric discontinuities that can markedly influence crack growth and ultimate failure \cite{Khosravani2020}. Within this broader context, Brighenti et al. examined the influence of DLP process parameters on both tensile and fracture properties of printed photopolymer specimens, directly demonstrating the sensitivity of fracture-related metrics to exposure and layer-thickness settings \cite{Brighenti2023}. Lenti further demonstrated the utility of digital image correlation (DIC) for fracture-toughness assessment in additively manufactured specimens \cite{Lenti2019}, while Sánchez et al. showed in a related AM polymer system that fracture-based criteria provide a robust framework for rationalizing failure in notched printed parts \cite{Sanchez2022}. Additional recent stereolithography-based fracture studies employing DIC and mechanics-based analysis further reinforce the relevance of fracture-sensitive characterization for photo-cured AM polymers \cite{Khosravani2023,Khosravani2024}. Taken together, these studies highlight fracture resistance as a physically meaningful and practically consequential target property for brittle or glassy DLP-printed photopolymers.

Despite these important prior efforts, the effect of DLP processing conditions on fracture resistance remains inadequately understood. Most available studies probe only the individual or partially coupled influence of a small number of processing variables, often at a limited set of discrete settings selected for practical convenience rather than to fully resolve the underlying process–property relationship. For example, studies commonly compare only a few layer orientations, a few layer thicknesses, or a narrow set of curing conditions \cite{Farkas2023,Jiang2023,Lee2018,Brighenti2023,Tang2022}, while related investigations on filament-based additive manufacturing methods likewise tend to focus on selected print orientations, infill levels, or thermal settings for a fixed specimen class or geometry \cite{Villacres2020,Upadhyay2017,Cantrell2017,Aourik2021,Hosseinzadeh2022}. Such approaches are useful for identifying broad trends, but they do not rigorously quantify the potentially complex interaction effects among processing variables that may govern mechanical performance in a multidimensional parameter space. Indeed, prior studies already suggest that different printing variables can interact in nontrivial ways through cure behavior, viscosity changes, interlayer bonding, and architecture-dependent load transfer. The practical difficulty is straightforward: even a modest four-parameter design space with five candidate values per parameter would require $5^4 = 625$ experiments in a full-factorial plan. As a result, the current literature has not yet established a continuous and sufficiently accurate functional relationship linking DLP processing parameters to fracture behavior in printed photopolymer parts.

Bayesian active learning offers an attractive path forward for this class of problem. In broad terms, Bayesian-active-learning-guided experimental design uses information extracted from prior experiments and/or simulations to construct a probabilistic surrogate of the response surface and then sequentially selects the next experiment according to a principled acquisition strategy \cite{Greenhill2020,Roussel2021}. In materials science and mechanics, Gaussian process regression (GPR)-based active learning and Bayesian optimization have emerged as powerful approaches for navigating expensive, noisy, and high-dimensional design spaces while balancing exploration of uncertain regions against exploitation of promising regions \cite{Imani2020,Liang2021,Lookman2019,Kusne2020,Bock2019}. These methods can substantially reduce the number of experiments required to identify high-performing conditions, while simultaneously learning the underlying process--property relationship and providing uncertainty estimates that are difficult to obtain through conventional one-factor-at-a-time or coarse factorial approaches. A further advantage is that once a reliable surrogate has been learned, it can be interrogated for interpretability and decision support. This can include one-at-a-time surrogate-based sensitivity analysis around the design space center to directly visualize how the response varies with each processing parameter, as well as variance-based global sensitivity analysis such as Sobol methods to quantify the relative contribution of each input and its interactions \cite{Morris1991,Ghanem2017,Marrel2009,Chauhan2024,Upadhyay2024}. In this sense, Bayesian active learning is valuable not only as a search strategy, but also as a framework for building data-efficient and interpretable maps of multivariable processing spaces.

Motivated by the incomplete understanding of process--fracture relationships in DLP printing and by the emerging opportunities afforded by Bayesian active learning, the present study demonstrates a machine-learning-guided exploration of fracture resistance in a DLP-printed model photopolymer \cite{Li2019}. Specifically, we combine DIC-assisted fracture characterization in three-point bending with a Bayesian active learning framework based on GPR and a modified upper confidence bound acquisition function to guide sequential exploration of the DLP parameter space. The resulting surrogate model is then used to identify high-performing regions within the investigated process window, construct a continuous process--fracture map, perform one-at-a-time surrogate-based sensitivity analysis around the design space center, and conduct Sobol-based global sensitivity analysis to rank the relative influence of key printing variables. Relative to prior studies that rely primarily on sparse discrete parameter comparisons, the present approach seeks to extract substantially richer process--property information from a modest experimental campaign while preserving physical interpretability.

The remainder of this paper is organized as follows. Section \ref{sec:Materials_and_Methods} describes the material system, specimen fabrication, fracture testing procedure, DIC-based response extraction, Bayesian active learning framework, and surrogate-based sensitivity analyses. Section \ref{sec:results_and_discussion} presents the experimentally measured fracture dataset, the learned process--fracture relationships, the one-at-a-time surrogate-based response trends, and the associated global sensitivity analysis, followed by a discussion of the governing physical trends and practical implications. Finally, Section \ref{sec:summary_and_conclusion} summarizes the main findings and outlines the broader significance of the proposed framework for data-efficient process mapping of DLP-printed photopolymers.

\section{Materials and Methods}
\label{sec:Materials_and_Methods}

\subsection{Material system, specimen fabrication, and process parameters}
\label{subsec:Material}

The model material system investigated in this study is a photosensitive, DLP-printable photopolymer recently developed by Li et al. \cite{Li2019}. This material was selected as a representative high-performance DLP resin because it combines good printability, relatively high cured-state stiffness and strength, and added functional attributes such as shape recovery and recyclability. The resin was synthesized following the preparation method of Li et al. \cite{Li2019}, using bisphenol A glycerolate dimethacrylate as the base monomer together with 2-hydroxy-2-methylpropiophenone (3.5 wt.\%), triethylamine (0.5 wt.\%), 1,4-butanediol dimethacrylate (20 wt.\%), and phenylbis(2,4,6-trimethylbenzoyl) phosphine oxide (1 wt.\%). The constituents were stirred at 72 $^{\circ}$C for 4 h under light-protected conditions, after which the resin was filtered and stored in an opaque container to minimize unintended ambient curing prior to printing. The resulting polymer has a glass transition temperature of approximately 95 $^{\circ}$C \cite{Li2019}.

\begin{figure}[b!]
    \centering
    \includegraphics[width=6.206in]{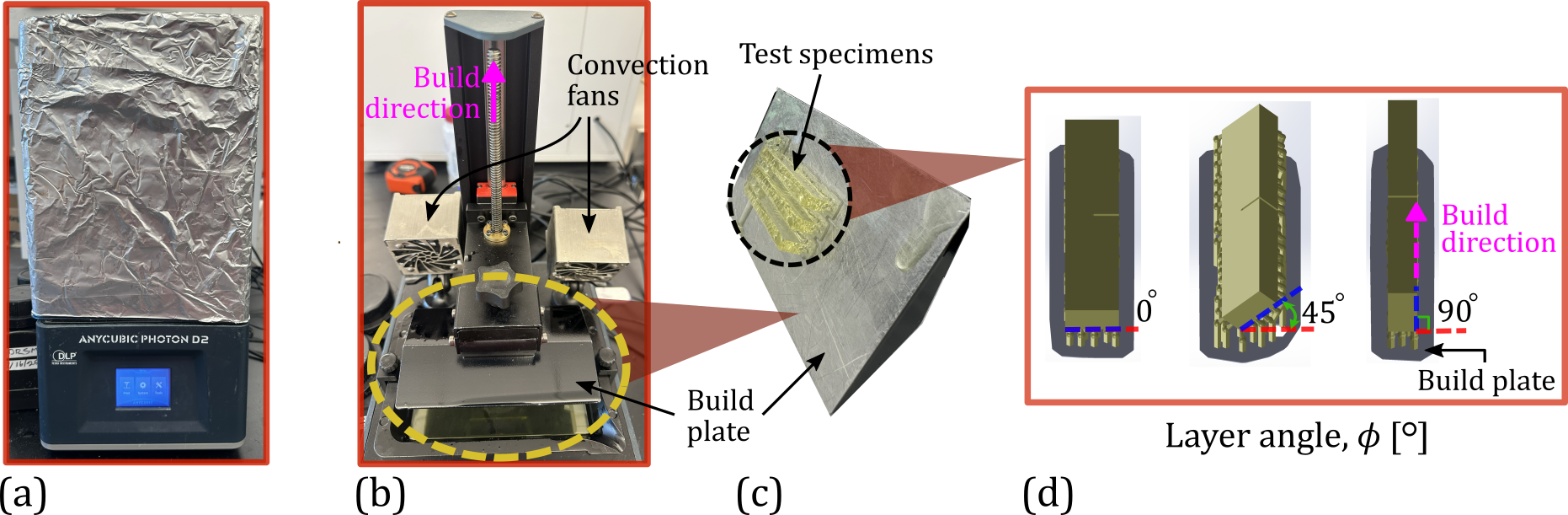}
    \caption{%
    DLP fabrication setup and specimen design used in the present study: (a) external view of the Anycubic Photon D2 printer during specimen fabrication, with ambient-light shielding in place; (b) open-printer view of the realized fabrication setup, showing the build plate and external convection fans used for print-temperature control; (c) three replicate fracture specimens printed in a single build and shown attached to the removed build plate; and (d) CAD representations of the fracture specimen illustrating the definition of the layer angle $\phi$ relative to the build plate and build direction.
    }
    \label{fig:DLP_printer}
\end{figure}

All specimens used in the present experimental campaign were fabricated using an Anycubic Photon D2 DLP printer (Anycubic, Inc., USA). In the realized setup, shown in Fig.~\ref{fig:DLP_printer}, external convection fans were used to control the printing temperature environment, and the printer assembly was shielded from ambient light during fabrication because of the strong photosensitivity of the resin. The process--fracture study focused on four printing parameters that could be varied independently during fabrication: layer angle $\phi$ (see Fig.~\ref{fig:DLP_printer}(d)), UV exposure time $t_{UV}$, layer height $h$, and print temperature $T$. The investigated ranges are summarized in Table~\ref{tab:processing_parameters}. Briefly, $\phi$ was varied from 0$^\circ$ to 90$^\circ$ with a resolution of 1$^\circ$, $t_{UV}$ from 7.0 to 12.0 s with a resolution of 0.5 s, $h$ from 50 to 150 $\mu$m with a resolution of 10 $\mu$m, and $T$ from 25 to 45 $^{\circ}$C with a resolution of 1 $^{\circ}$C. Here, the layer angle $\phi$ denotes the angle between the plane of the broad specimen face and the build plate (equivalently, the printed layer plane), such that $\phi=0^\circ$ corresponds to the specimen plane being parallel to the build plate and normal to the build direction, whereas $\phi=90^\circ$ corresponds to the specimen plane containing the build direction. The print temperature denotes the local thermal environment imposed during printing using the convection-fan arrangement, whereas UV exposure time denotes the duration for which each layer was exposed to UV light during curing. For reference, the UV irradiance reported for the Photon D2 is approximately 2.92~$\mathrm{mW/cm^2}$ at 405~nm \cite{Liqcreate_UV_Intensity}. These parameter bounds were selected based on the operating limits and practical stability of the printing setup. In particular, the lower bound on UV exposure time was constrained by print reliability, while the temperature range was also influenced by resin viscosity and reflow behavior during layer recoating. Within each build, all processing variables other than the four parameters of interest were held fixed as far as practically possible. After printing, the specimens were cleaned with 70\% isopropyl alcohol and lightly sanded using successive 240- and 600-grit sandpapers to remove residual resin and support material while minimizing unintended geometric alteration.

Fracture specimens were designed in Fusion 360 (Autodesk, Inc., USA) and printed as single-edge-notched bending (SENB)-type coupons for subsequent three-point bending tests, following the general SENB geometry recommended in ASTM D5045 \cite{ASTMD5045} (see Fig. \ref{fig:fracture_testing}(b) for dimensions). The printed specimens had nominal dimensions of 50 mm × 10 mm × 5 mm. A notch of width 0.5 mm was incorporated into the geometry and extended 5 mm upward from the 50 mm × 10 mm face. For each parameter combination, three nominally identical specimens were printed together in a single build to assess repeatability at that setting, as shown in Fig.~\ref{fig:DLP_printer}(c), consistent with the ASTM recommendation of at least three replicate tests per material condition. Because the printer resolution was insufficient to produce a sufficiently sharp notch tip directly, a starter crack was introduced after printing using a blade-assisted initiation step, in keeping with the notch-sharpening and razor-precracking guidance of ASTM D5045 for SENB specimens. Specifically, a custom fixture mounted on a universal testing machine (Model 240, TestResources, Inc., USA) equipped with a 1.1 kN load cell was used to press a sharp blade into the printed notch until an applied load of approximately 15 N was reached. This initiation load was selected empirically based on preliminary trials to produce a sharp crack tip while minimizing premature crack growth, and each specimen was subsequently inspected visually to verify that no unintended crack propagation had occurred prior to fracture testing.

\begin{table}[t!]
\centering
\caption{DLP processing parameters considered in the present study.}
\label{tab:processing_parameters}
\begin{tabular}{lcc}
\toprule
\small{\textbf{Processing parameter}} & \small{\textbf{Range [min, max]}} & \small{\textbf{Resolution}} \\
\midrule
\small{Layer angle, $\phi$}         & \small{$[0^\circ,\,90^\circ]$}              & \small{$1^\circ$} \\
\small{UV exposure time, $t_{UV}$}  & \small{$[7.0~\mathrm{s},\,12.0~\mathrm{s}]$} & \small{$0.5~\mathrm{s}$} \\
\small{Layer height, $h$}           & \small{$[50~\mu\mathrm{m},\,150~\mu\mathrm{m}]$} & \small{$10~\mu\mathrm{m}$} \\
\small{Print temperature, $T$}      & \small{$[25^\circ\mathrm{C},\,45^\circ\mathrm{C}]$} & \small{$1^\circ\mathrm{C}$} \\
\bottomrule
\end{tabular}
\end{table}


\subsection{Fracture testing and DIC-based response extraction}
\label{subsection:DIC_J_c}

\begin{figure}[t!]
    \centering
    \includegraphics[width=5.911in]{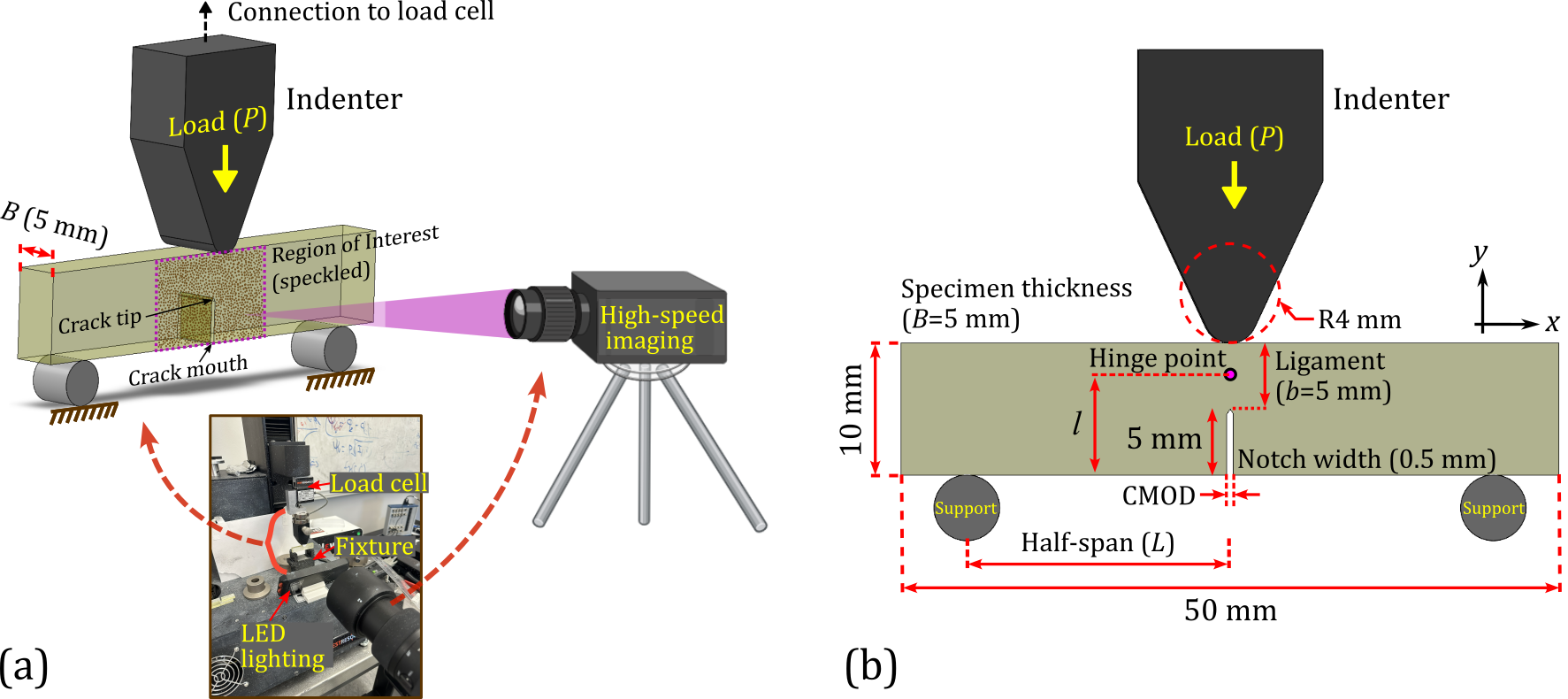}
    \caption{%
    Fracture testing and J-integral evaluation framework: (a) displacement-controlled three-point bending setup with high-speed imaging for digital image correlation (DIC), showing the indenter connected to the load cell, support fixture, LED illumination, speckled region of interest, crack tip, and crack mouth; and (b) idealized single-edge-notched bending (SENB) configuration showing the indenter radius, specimen geometry, and quantities used in the J-integral calculation, including the applied load $P$, specimen thickness $B$, remaining ligament length $b$, crack mouth opening displacement ($\mathrm{CMOD}$), crack-mouth-to-hinge-point distance $l$, and half-span $L$.
    }
    \label{fig:fracture_testing}
\end{figure}

Fracture resistance was characterized using displacement-controlled three-point bending tests on the printed SENB-type specimens described in the previous subsection. Figure~\ref{fig:fracture_testing}(a) shows the test setup, in which each specimen was supported in a custom three-point bending fixture and loaded in Mode I by a wedge-shaped indenter connected to the universal testing machine. The indenter was driven at a constant crosshead displacement rate, while the applied load $P$ was measured using the 1.1~kN load cell. Simultaneously, the deformation field in the crack-tip region was recorded using high-speed imaging for subsequent digital image correlation (DIC). The speckled specimen face was aligned parallel to the camera image plane, with the camera optical axis approximately normal to the specimen surface, and an LED light source was arranged to maximize the contrast of the speckle pattern. A Phantom Miro M110 high-speed camera (Vision Research, Inc., USA) was positioned approximately 30~cm from the specimen. Based on preliminary trials, a frame rate of 20,000~fps was selected as a compromise between temporal resolution and recording duration: this rate was sufficiently high to resolve crack initiation and early propagation over multiple frames while retaining a sufficiently long acquisition window to capture the loading history from the beginning of the test through fracture initiation. Because of storage limitations at this frame rate, the videos were acquired at a resolution of $192\times320$ pixels, providing a maximum recording duration of approximately 1.7~s. The optical magnification was approximately 1:1.7 over a region of interest of approximately $10~\mathrm{mm}\times10~\mathrm{mm}$. The tests were conducted at a nominal loading rate of $2\times10^{-1}~\mathrm{s}^{-1}$, defined here as the crosshead velocity divided by the 10~mm specimen depth.

To enable high-speed DIC, a high-contrast random speckle pattern was applied uniformly to the specimen surface using white airbrush paint, with particular attention to the crack-tip region and crack mouth. The captured image sequences were processed in a commercial DIC software package (MatchID US, Inc.) using a two-dimensional DIC workflow. For each test, the image set was calibrated once using the known specimen depth to convert pixel coordinates to physical dimensions. A rectangular region of interest was then defined over the remaining ligament above the crack tip. In addition, a virtual extensometer was placed across the crack mouth to measure the crack mouth opening displacement ($\mathrm{CMOD}$) as a function of time. Green--Lagrange strain was used for strain evaluation. Following general DIC recommendations \cite{Jones2025}, subset and step sizes were selected based on the observed speckle size, with the subset size taken to be approximately three times the average speckle size and the step size approximately one-half of the subset size. A larger subset was used for the crack-mouth extensometer to reduce noise in the $\mathrm{CMOD}$ signal.

The DIC measurements were used to aid computation of the J-integral history during each test. The J-integral represents the energy available for crack extension per unit newly created crack area and therefore has units of energy per unit area, $\mathrm{J/m^2}$, equivalently force per unit length, $\mathrm{N/m}$. In the present work, $J$ is reported in $\mathrm{N/mm}$, which is numerically equivalent to $\mathrm{kJ/m^2}$. For a deeply notched specimen subjected to bending, as represented schematically in Fig.~\ref{fig:fracture_testing}(b), Rice et al.~\cite{Rice1973} showed that the J-integral can be expressed in terms of the applied moment and the corresponding rotation of the remaining uncracked ligament as
\begin{equation}
\label{eq:J_integral}
    J = \frac{2}{Bb}\int_0^\theta Md\theta,
\end{equation}
where $B$ is the specimen thickness, $b$ is the remaining ligament length, $M$ is the bending moment, and $\theta$ is the ligament rotation. In the present configuration, the hinge point was defined as the location of zero axial strain within the ligament above the crack tip, separating the compressive region above from the tensile region below. This point was identified from the DIC-derived horizontal strain field, $E_{xx}$, by extracting the axial strain along a vertical line extending from the crack tip toward the upper specimen surface and locating the zero-strain crossing. To improve robustness against frame-to-frame noise, the hinge-point location was averaged over the 50 frames immediately preceding visible crack propagation.

Using the small-rotation approximation, the ligament rotation was computed from
\begin{equation}
\label{eq:theta}
    \theta = \frac{\mathrm{CMOD}}{l},
\end{equation}
where $l$ is the distance from the crack mouth to the hinge point, as shown in Fig.~\ref{fig:fracture_testing}(b). The applied bending moment was obtained from statics of the three-point bending configuration as
\begin{equation}
\label{eq:moment}
    M = \frac{P}{2}L,
\end{equation}
where $P$ is the measured applied load and $L$ is the half-span defined in Fig.~\ref{fig:fracture_testing}(b). Using the time histories of $P$, $\mathrm{CMOD}$, and the hinge-point location, a MATLAB script was used to compute $M(t)$, $\theta(t)$, and the corresponding J-integral history for each specimen through numerical integration of Eq.~(\ref{eq:J_integral}) using the trapezoidal rule. In a typical test, the $\mathrm{CMOD}$ initially increased gradually, accelerated as fracture initiation approached, and then exhibited an abrupt jump upon unstable crack propagation. The corresponding $J$ history displayed an approximately sigmoidal evolution, with a gradual initial increase followed by a more rapid rise and subsequent leveling toward a peak or near-asymptotic value immediately before visible crack propagation. This critical value, denoted by $J_c$, was taken as the fracture resistance metric for an individual specimen. The response associated with each printing condition was defined as the average $J_c$ across the corresponding replicate specimens.

\subsection{Bayesian-active-learning-guided experimental design and process--fracture mapping}
\label{sec:bayesian_active_learning}

A major obstacle to exploring high-dimensional process-design spaces is the prohibitively large number of experiments required for exhaustive testing. In the present study, the DLP processing space is defined by the four input variables $\{\phi,t_{UV},h,T\}$ listed in Table~\ref{tab:processing_parameters}. Considering the discretization adopted here, this corresponds to 91 possible layer angles $\phi$ ($0^\circ$--$90^\circ$), 11 possible UV exposure times $t_{UV}$ (7.0--12.0~s), 11 possible layer heights $h$ (50--150~$\mu$m), and 21 possible print temperatures $T$ (25--45~$^\circ$C). A traditional full-factorial search of this space would therefore require $91\times 11\times 11\times 21 = 231{,}231$ experiments, which is clearly impractical for a fracture-testing campaign. To address this challenge, a Bayesian-active-learning-based closed-loop experimental design was developed to efficiently explore the process space using a relatively small number of experiments.

\begin{algorithm}[t!]
\caption{Bayesian-active-learning-based closed-loop experimental design for process--fracture mapping}
\label{alg:bayesian_active_learning}
\begin{algorithmic}[1]
\Require Physical discrete design space $\Omega$ defined by Table~\ref{tab:processing_parameters}; min--max transformation in Eq.~(\ref{eq:minmax_scaling}); initial number of random experiments $n_0=2$; total number of active learning iterations $N_{\mathrm{iter}}$
\Ensure Final dataset $\mathcal{D}$ and trained GPR surrogate model $\mathcal{M}$

\State Construct the normalized discrete design space $\tilde{\Omega}_d$ by applying Eq.~(\ref{eq:minmax_scaling}) to all admissible processing combinations in $\Omega$
\State Randomly select two initial physical processing conditions, $\mathbf{X}_0,\mathbf{X}_1\in\Omega$

\For{$t=0,1$}
    \State Fabricate specimens using physical processing condition $\mathbf{X}_t$
    \State Perform three-point bending fracture tests and compute replicate-averaged critical J-integral $J_{c,t}$
    \State Normalize $\mathbf{X}_t$ using Eq.~(\ref{eq:minmax_scaling}) to obtain $\tilde{\mathbf{X}}_t$
\EndFor

\State Construct initial normalized training dataset:
\[
\mathcal{D}_2=\{(\tilde{\mathbf{X}}_t,J_{c,t})\}_{t=0}^{1}
\]

\For{$i=1$ to $N_{\mathrm{iter}}$}
    \State Train GPR surrogate model $\mathcal{M}_i$ on current normalized dataset $\mathcal{D}_{i+1}$

    \State Compute the randomized trade-off coefficient:
    \[
    \kappa_i=
    \frac{\log\!\left[\left(1/\sqrt{2\pi}\right)\left((i+1)^2+1\right)\right]}
    {\log(1+\vartheta/2)},
    \qquad
    \beta_i \sim \Gamma(\kappa_i,\vartheta)
    \]

    \State Define the variance-amplified optimistic acquisition function:
    \[
    \mathcal{A}_i(\tilde{\mathbf{X}})
    =
    \mu_i(\tilde{\mathbf{X}})
    +
    \sqrt{\beta_i}\,\sigma_i^2(\tilde{\mathbf{X}})
    \]

    \State Select the next normalized processing condition from the admissible grid:
    \[
    \tilde{\mathbf{X}}_{i+1}
    =
    \argmax_{\tilde{\mathbf{X}}\in\tilde{\Omega}_d}
    \mathcal{A}_i(\tilde{\mathbf{X}})
    \]

    \State Map $\tilde{\mathbf{X}}_{i+1}$ back to the physical processing condition $\mathbf{X}_{i+1}\in\Omega$ using Eq.~(\ref{eq:inverse_minmax})

    \State Fabricate specimens using $\mathbf{X}_{i+1}$
    \State Perform three-point bending fracture tests and compute replicate-averaged critical J-integral $J_{c,i+1}$

    \State Augment the normalized dataset:
    \[
    \mathcal{D}_{i+2}
    =
    \mathcal{D}_{i+1}
    \cup
    \{(\tilde{\mathbf{X}}_{i+1},J_{c,i+1})\}
    \]
\EndFor

\State Train the final GPR surrogate model $\mathcal{M}$ on the full normalized dataset $\mathcal{D}$
\State Perform leave-one-out cross-validation to assess surrogate reliability
\State Use the final surrogate $\mathcal{M}$ for one-at-a-time and global sensitivity analyses

\end{algorithmic}
\end{algorithm}

The overall workflow is summarized in Algorithm~\ref{alg:bayesian_active_learning}. The procedure was initialized by fabricating and testing specimens corresponding to two randomly selected combinations of DLP processing conditions, denoted by $\mathbf{X}_0=\{\phi_0,t_{UV,0},h_0,T_0\}$ and $\mathbf{X}_1=\{\phi_1,t_{UV,1},h_1,T_1\}$. For each processing condition, the fracture response was quantified by the replicate-averaged critical J-integral, denoted here by $J_{c,0}$ and $J_{c,1}$. Because the four input variables have different physical units and ranges, the physical processing vectors were normalized before surrogate modeling using a min--max transformation,
\begin{equation}
\label{eq:minmax_scaling}
\tilde{X}_j=
\frac{X_j-X_j^{\min}}
{X_j^{\max}-X_j^{\min}},
\qquad j=1,\dots,d,
\end{equation}
where $X_j^{\min}$ and $X_j^{\max}$ denote the lower and upper bounds of the $j$-th processing parameter listed in Table~\ref{tab:processing_parameters}, $d=4$ is the input dimensionality, and the input ordering is $X_1=\phi$, $X_2=t_{UV}$, $X_3=h$, and $X_4=T$. Thus, each physical input vector $\mathbf{X}=[\phi,\,t_{UV},\,h,\,T]$ was mapped to a normalized input vector $\tilde{\mathbf{X}}=[\tilde{\phi},\,\tilde{t}_{UV},\,\tilde{h},\,\tilde{T}]\in[0,1]^4$. These initial experiments and the normalization pre-processing generated the first input--output dataset,
\begin{equation}
\mathcal{D}_2=\{(\tilde{\mathbf{X}}_t,J_{c,t})\}_{t=0}^{1},
\end{equation}
on which a Gaussian process regression (GPR) surrogate model was trained.

In the present context, GPR provides a nonparametric probabilistic regression framework that learns the statistical relationship between the four-dimensional processing-condition vector and the resulting fracture resistance, while also quantifying prediction uncertainty throughout the admissible design space. For details on the statistical foundations of GPR, the reader is referred to Williams and Rasmussen \cite{Williams1995}. The surrogate model is written as
\begin{equation}
\label{eq:gpr_map}
\mathcal{M}:\tilde{\mathbf{X}}\mapsto J_c,
\end{equation}
where $\tilde{\mathbf{X}}\in[0,1]^4$ is the normalized input processing vector and $J_c\in\mathbb{R}$ is the corresponding replicate-averaged critical J-integral. In GPR, the latent response function is assumed to follow a Gaussian process (GP) prior,
\begin{equation}
\mathcal{M}(\tilde{\mathbf{X}})\sim\mathcal{GP}\!\left(0,K(\tilde{\mathbf{X}},\tilde{\mathbf{X}}')\right),
\end{equation}
where $K(\tilde{\mathbf{X}},\tilde{\mathbf{X}}')$ is the covariance kernel. In this work, a Mat\'ern-$3/2$ kernel was adopted because of its flexibility in representing moderately smooth nonlinear response surfaces while retaining robustness for sparse datasets \cite{Matern1986,Thant2025,Stein1999}. For two normalized input vectors $\tilde{\mathbf{X}}_p$ and $\tilde{\mathbf{X}}_q$, the Mat\'ern-$3/2$ covariance is written as
\begin{equation}
\label{eq:matern_kernel}
k_{\mathrm{M32}}(\tilde{\mathbf{X}}_p,\tilde{\mathbf{X}}_q)
=
\left(
1+\frac{\sqrt{3}r_{pq}}{\ell_{\mathrm{GP}}}
\right)
\exp\!\left(
-\frac{\sqrt{3}r_{pq}}{\ell_{\mathrm{GP}}}
\right),
\end{equation}
where $r_{pq}=\|\tilde{\mathbf{X}}_p-\tilde{\mathbf{X}}_q\|_2$ is the Euclidean distance in normalized input space and $\ell_{\mathrm{GP}}$ is the isotropic kernel lengthscale. During training, a nugget parameter $\alpha$ was added to the diagonal of the covariance matrix,
\begin{equation}
K_{pq}
=
k_{\mathrm{M32}}(\tilde{\mathbf{X}}_p,\tilde{\mathbf{X}}_q)
+
\alpha\,\delta_{pq},
\end{equation}
where $\delta_{pq}$ is the Kronecker delta. The nugget parameter improves numerical conditioning of the covariance matrix and can also be interpreted as the variance of additional independent Gaussian measurement noise in the training observations. Even when trained on the small initial dataset $\mathcal{D}_2$, the GPR model provides a posterior mean prediction $\mu(\tilde{\mathbf{X}})$ and associated variance $\sigma^2(\tilde{\mathbf{X}})$ across the entire normalized processing domain. These two quantities are then used to define an acquisition function for selecting the next experiment.

The next measurement point was chosen using a modified upper-confidence-bound (UCB) \cite{Auer2003,Srinivas2009}-style acquisition function of the form
\begin{equation}
\label{eq:ucb_modified}
\mathcal{A}(\tilde{\mathbf{X}})=\mu(\tilde{\mathbf{X}})+\sqrt{\beta}\,\sigma^2(\tilde{\mathbf{X}}),
\end{equation}
where $\beta$ is a trade-off coefficient controlling exploration versus exploitation. Unlike the classical GP-UCB acquisition \cite{Bian2021}, which weights the posterior standard deviation, Eq.~(\ref{eq:ucb_modified}) employs the posterior variance directly and therefore places stronger emphasis on highly uncertain regions of the process space. In this sense, the adopted rule may be viewed as a variance-amplified optimistic acquisition function designed to promote exploration in a sparse-data, experimentally expensive setting. This choice is also broadly consistent with the growing use of variance-aware criteria in Bayesian optimization, where variance is treated explicitly rather than only through the standard-deviation term \cite{Makarova2021}. In the present work, this modified acquisition was adopted as implemented throughout the experimental campaign and was retained consistently for all reported results.

The next normalized processing condition to test is then obtained by solving
\begin{equation}
\label{eq:next_point}
\tilde{\mathbf{X}}_{i+1}
=
\argmax_{\tilde{\mathbf{X}}\in\tilde{\Omega}_d}
\mathcal{A}(\tilde{\mathbf{X}}),
\end{equation}
where $\tilde{\Omega}_d$ denotes the normalized discrete design space obtained by applying Eq.~(\ref{eq:minmax_scaling}) to all experimentally admissible parameter combinations listed in Table~\ref{tab:processing_parameters}. Because the acquisition function was maximized only over this discrete admissible grid, the selected point corresponded directly to an experimentally realizable processing condition. The corresponding physical processing condition, $\mathbf{X}_{i+1}$, was obtained by inverse-normalizing $\tilde{\mathbf{X}}_{i+1}$ according to
\begin{equation}
\label{eq:inverse_minmax}
X_{j,i+1}
=
X_j^{\min}
+
\tilde{X}_{j,i+1}
\left(
X_j^{\max}-X_j^{\min}
\right),
\qquad j=1,\dots,d.
\end{equation}
After the corresponding experiment was performed and the new $J_c$ value was measured, the normalized input--output pair was appended to the dataset, the GPR model was retrained, and the acquisition function was updated. Repetition of this loop yields a statistically guided experimental campaign in which each new experiment is selected to improve knowledge of the process--fracture map while also favoring high-performing regions of the design space.

To balance broad exploration of the design space against efficient movement toward high-$J_c$ regions, a randomized trade-off strategy inspired by the randomized Gaussian process upper confidence bound (RGP-UCB) method of Berk et al.~\cite{Berk2021} was employed. Rather than fixing the trade-off coefficient $\beta$, this approach updates it stochastically at each iteration by drawing from a Gamma distribution,
\begin{equation}
\beta_i\sim\Gamma(\kappa_i,\vartheta),
\end{equation}
where $\kappa_i$ is the iteration-dependent shape parameter and $\vartheta$ is a fixed scale parameter. Following Berk et al.~\cite{Berk2021}, the shape parameter was updated as
\begin{equation}
\label{eq:kappa}
\kappa_i=
\frac{\log\!\left[\left(1/\sqrt{2\pi}\right)\left((i+1)^2+1\right)\right]}
{\log(1+\vartheta/2)},
\end{equation}
and $\vartheta=1$ was used throughout the present work. In contrast to a fixed-$\beta$ policy, this randomized strategy helps prevent premature convergence to local optima and promotes a more balanced exploration of the process space as the campaign progresses.

At the conclusion of the experimental campaign, the accumulated dataset and the final trained GPR model together define a continuous statistical process--fracture map linking DLP processing conditions to the critical J-integral. The reliability of this surrogate model was assessed using leave-one-out cross-validation (LOOCV). In this procedure, each experimental datapoint was removed in turn, the GPR model was retrained on the remaining data, and the withheld response was predicted. Model reliability was then quantified using the coefficient of determination ($R^2$) and Pearson correlation coefficient ($r$) between the predicted and experimentally measured $J_c$ values over all LOOCV folds.

\subsection{Surrogate-based sensitivity analysis}
\label{subsection:surrogate_sensitivity}

The final trained GPR surrogate was interrogated using both one-at-a-time surrogate-based sensitivity analysis around the design space center and variance-based global sensitivity analysis. The former provides direct visualization and local metrics describing how the predicted fracture resistance varies with each individual processing parameter under fixed conditions, whereas the latter quantifies the relative contribution of each parameter, including interaction effects, to the overall variance in the surrogate-predicted response.

First, a one-at-a-time surrogate-based sensitivity analysis \cite{Ghanem2017} was performed around the center of the normalized design space. Because the GPR model was trained using the normalized input vector $\tilde{\mathbf{X}}$, as described in Section~\ref{sec:bayesian_active_learning}, the design space center is given by
\begin{equation}
\tilde{\mathbf{X}}_c =
[0.5,\,0.5,\,0.5,\,0.5].
\end{equation}
In physical coordinates, this point corresponds to $\phi=45^\circ$, $t_{UV}=9.5~\mathrm{s}$, $h=100~\mu\mathrm{m}$, and $T=35^\circ\mathrm{C}$. For each processing parameter, a one-dimensional conditional response curve was generated by varying that parameter across its normalized range while holding all other parameters fixed at their center values. Specifically, for the $j$-th processing parameter, we define the conditional input vector
\begin{equation}
\tilde{\mathbf{X}}^{(j)}(\tilde{x}_j)
=
\left[
\tilde{X}_1^{(j)},\ldots,\tilde{X}_d^{(j)}
\right],
\end{equation}
with components
\begin{equation}
\tilde{X}_k^{(j)}
=
\begin{cases}
\tilde{x}_j, & k=j,\\
0.5, & k\neq j,
\end{cases}
\qquad k=1,\ldots,d.
\end{equation}
The corresponding local response function was then defined as
\begin{equation}
g_j(\tilde{x}_j)
=
\mathcal{M}\!\left(\tilde{\mathbf{X}}^{(j)}(\tilde{x}_j)\right),
\qquad \tilde{x}_j\in[0,1],
\end{equation}
where $j=1,\dots,d$ and $d=4$. Each one-dimensional response curve was generated from the posterior mean of the trained GPR model evaluated at $N_{\mathrm{local}}=1000$ uniformly spaced points over $\tilde{x}_j\in[0,1]$. The resulting curve was then plotted against the corresponding physical processing variable obtained through the inverse min--max transformation in Eq.~(\ref{eq:inverse_minmax}). Thus, the curves provide a direct representation of the learned dependence of $J_c$ on each processing variable around the center of the design space. For each parameter, the local response range was also computed as
\begin{equation}
\label{eq:local_range}
\Delta J_j
=
\max_{\tilde{x}_j\in[0,1]} g_j(\tilde{x}_j)
-
\min_{\tilde{x}_j\in[0,1]} g_j(\tilde{x}_j),
\end{equation}
which quantifies the total variation in the surrogate-predicted $J_c$ when parameter $X_j$ is varied across its admissible range while all other parameters are fixed at the design space center.

To compare the local influence of different parameters, local slopes were computed from the one-at-a-time response curves. Since the GPR surrogate was trained in normalized coordinates, the primary local sensitivity metric was defined with respect to the normalized input variable,
\begin{equation}
\label{eq:local_sensitivity_normalized}
m_j=
\left.
\frac{d g_j}{d\tilde{x}_j}
\right|_{\tilde{x}_j=0.5},
\end{equation}
and was estimated using a central finite-difference approximation. This normalized local sensitivity has units of N/mm and provides a consistent basis for comparing the relative influence of different processing parameters, despite their different physical units and ranges. For interpretability in physical coordinates, an additional physical-coordinate local sensitivity was computed as
\begin{equation}
\label{eq:local_sensitivity_physical}
m_j^{\mathrm{phys}}
=
\left.
\frac{d g_j}{d x_j}
\right|_{x_j=x_j^c}
=
\frac{m_j}{X_j^{\max}-X_j^{\min}},
\end{equation}
where $x_j^c$ is the physical value of parameter $X_j$ at the design space center. The three local quantities $\Delta J_j$, $m_j$, and $m_j^{\mathrm{phys}}$ therefore provide complementary information: $\Delta J_j$ measures the total predicted variation across the investigated range, $m_j$ enables normalized cross-parameter comparison of the local sensitivity slope at the design space center, and $m_j^{\mathrm{phys}}$ expresses the local slope in the physical units of each processing parameter.

The final GPR surrogate was further used for global sensitivity analysis to quantify the relative influence of the four processing parameters on fracture resistance over the full investigated process window. Variance-based Sobol analysis \cite{Sobol1993,Sobol2001} was adopted for this purpose because it decomposes the variance of the model output into contributions from individual inputs and their interactions. Since the GPR surrogate was trained in normalized input space, the Sobol analysis was performed using the normalized input vector $\tilde{\mathbf{X}}$. Letting $Y=\mathcal{M}(\tilde{\mathbf{X}})$ denote the surrogate-predicted fracture response, and using $\mathbb{E}[\cdot]$ and $\mathrm{Var}(\cdot)$ to denote expectation and variance operators, respectively, the first-order Sobol index for normalized input $\tilde{X}_j$ is defined as
\begin{equation}
\label{eq:Sobol_first-order}
S_j=
\frac{\mathrm{Var}_{\tilde{X}_j}\!\left(\mathbb{E}[Y\,|\,\tilde{X}_j]\right)}
{\mathrm{Var}(Y)},
\end{equation}
where $j=1,\dots,d$ indexes the input parameters, and $d=4$. Through the min--max transformation in Eq.~(\ref{eq:minmax_scaling}), these normalized inputs correspond directly to the physical processing parameters $\phi$, $t_{UV}$, $h$, and $T$. The first-order index $S_j$ measures the fraction of output variance attributable to the direct effect of input parameter $j$ alone. In general, the sum of the first-order indices satisfies $\sum_{j=1}^{d} S_j \le 1$, with any deficit from unity indicating the presence of interaction effects among the inputs. The corresponding total-order Sobol index is
\begin{equation}
\label{eq:Sobol_total-order}
S_j^{\mathrm{tot}}=
1-
\frac{
\mathrm{Var}_{\tilde{\mathbf{X}}_{\sim j}}\!\left(
\mathbb{E}[Y\,|\,\tilde{\mathbf{X}}_{\sim j}]
\right)
}
{\mathrm{Var}(Y)},
\end{equation}
where $\tilde{\mathbf{X}}_{\sim j}$ denotes the collection of all normalized input parameters except $\tilde{X}_j$. This index captures the full contribution of input parameter $j$, including both its direct effect and all interaction effects involving that parameter. Numerical estimation of these indices was performed using the Monte-Carlo-based Sobol pick-freeze scheme of Monod et al.~\cite{Monod2006}, together with Jansen's estimator for the total-order sensitivity index \cite{Jansen1999,Ghanem2017}. Full details of the numerical Sobol-index estimation are provided in \ref{app:sobol}. All surrogate modeling and sensitivity analysis calculations reported in this work were implemented in Python using standard scientific-computing libraries.

\section{Results and Discussion}
\label{sec:results_and_discussion}

The results are presented in three parts. First, we examine a representative fracture experiment to illustrate the measured deformation response and DIC-based extraction of the critical J-integral, $J_c$. Second, we present the Bayesian active learning campaign, including the explored DLP processing conditions and the predictive performance of the final GPR process--fracture surrogate. Finally, we use the trained surrogate model to interpret the learned process--fracture relationship through local one-at-a-time response curves and variance-based global sensitivity analysis.

\subsection{Representative fracture response and $J_c$ extraction}
\begin{figure}[t!]
    \centering
    \includegraphics[width=5.387in]{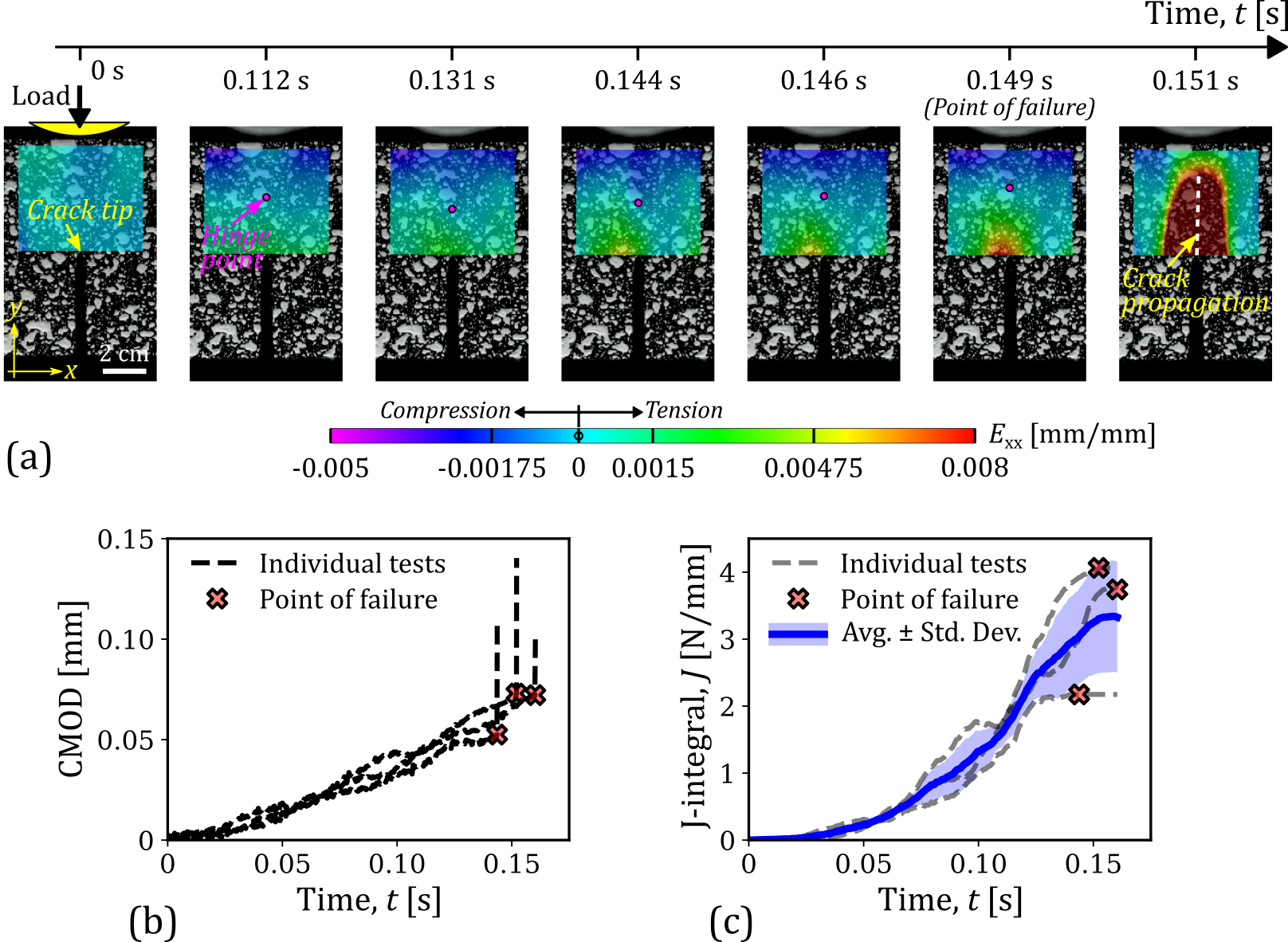}
    \caption{DIC-assisted extraction of the critical J-integral, $J_c$, from a representative fracture experiment conducted at $\phi=90^\circ$, $t_{UV}=12$~s, $h=150~\mu$m, and $T=45^\circ$C. (a) Selected snapshots showing the evolution of the axial Green--Lagrange strain field, $E_{xx}$, with the DIC-identified hinge point, point of failure, and subsequent Mode-I crack propagation indicated. The snapshots are concentrated toward the latter portion of the test, where the most pronounced strain localization occurs. (b) Crack mouth opening displacement (CMOD) histories for the three replicate specimens, with the point of failure marking the final pre-propagation datum. (c) Corresponding J-integral histories and their average $\pm$ one standard deviation. For visualization of the ensemble statistics, each individual J-integral history was held constant after failure. For each replicate, $J_c$ was taken as the J-integral value immediately preceding visible crack propagation, and the replicate-averaged value was used as the fracture resistance for that processing condition.}
    \label{fig:representative_fracture_response}
\end{figure}

We first examine a representative fracture experiment to illustrate the measured deformation response and extraction of the critical J-integral, $J_c$, following the DIC-assisted procedure described in Section~\ref{subsection:DIC_J_c} and used throughout the process--fracture mapping framework. The processing condition considered here was $\phi=90^\circ$, $t_{UV}=12$~s, $h=150~\mu$m, and $T=45^\circ$C, for which three nominally identical specimens were tested. Figure~\ref{fig:representative_fracture_response}(a) shows selected snapshots from one of these tests with the DIC-derived axial Green--Lagrange strain field, $E_{xx}$, overlaid on the specimen surface. The snapshots are intentionally non-uniformly spaced in time and emphasize the latter portion of the test, during which the most significant changes in the strain distribution occur. The bending-dominated deformation produces axial compression near the upper specimen surface beneath the indenter and axial tension in the remaining ligament near the crack tip. The zero-strain transition between these regions defines the hinge point introduced in Section~\ref{subsection:DIC_J_c}. As evident from the sequence, its instantaneous location varies slightly during loading; accordingly, the J-integral calculation used the hinge-point location averaged over the final 50 DIC frames preceding crack propagation. Meanwhile, tensile strain increasingly localizes near the crack tip as loading progresses. The frame at the identified point of failure represents the final frame before visible crack extension, whereas the subsequent frame shows rapid crack propagation approximately collinear with the starter notch, consistent with Mode-I fracture of the SENB specimen.

\begin{table}[t!]
\centering
\caption{Experimentally investigated DLP processing conditions and corresponding fracture resistance. The table lists the 28 distinct processing conditions evaluated during the Bayesian-active-learning-guided campaign (in the order they were tested). For each condition, the reported fracture resistance is the replicate-averaged critical J-integral, $J_c$.}
\label{tab:explored_conditions}
\begin{tabular}{cccccc}
\toprule
\small{\textbf{Trial no.}} & \multicolumn{4}{c}{\small{\textbf{Processing parameters}}} & \small{\textbf{Fracture resistance}} \\
\cmidrule(lr){2-5}
 & \small{$\phi$ [$^\circ$]} & \small{$t_{UV}$ [s]} & \small{$h$ [$\mu$m]} & \small{$T$ [$^\circ$C]} & \small{$J_c$ [N/mm]} \\
\midrule
\small{1}  & \small{0}  & \small{9.5}  & \small{100} & \small{35} & \small{3.6}  \\
\small{2}  & \small{90} & \small{9.0}  & \small{120} & \small{32} & \small{3.7}  \\
\small{3}  & \small{45} & \small{8.0}  & \small{80}  & \small{30} & \small{5.9}  \\
\small{4}  & \small{0}  & \small{7.0}  & \small{50}  & \small{25} & \small{12.8} \\
\small{5}  & \small{0}  & \small{9.5}  & \small{50}  & \small{25} & \small{7.0}  \\
\small{6}  & \small{5}  & \small{7.5}  & \small{50}  & \small{26} & \small{10.9} \\
\small{7}  & \small{0}  & \small{12.0} & \small{150} & \small{25} & \small{3.3}  \\
\small{8}  & \small{0}  & \small{12.0} & \small{150} & \small{45} & \small{2.6}  \\
\small{9}  & \small{90} & \small{12.0} & \small{50}  & \small{25} & \small{1.5}  \\
\small{10} & \small{0}  & \small{7.0}  & \small{50}  & \small{45} & \small{2.4}  \\
\small{11} & \small{90} & \small{7.0}  & \small{150} & \small{45} & \small{2.6}  \\
\small{12} & \small{90} & \small{12.0} & \small{50}  & \small{45} & \small{1.2}  \\
\small{13} & \small{90} & \small{12.0} & \small{150} & \small{25} & \small{1.5}  \\
\small{14} & \small{0}  & \small{12.0} & \small{50}  & \small{45} & \small{1.8}  \\
\small{15} & \small{90} & \small{7.0}  & \small{50}  & \small{25} & \small{3.6}  \\
\small{16} & \small{90} & \small{12.0} & \small{150} & \small{45} & \small{3.4}  \\
\small{17} & \small{0}  & \small{12.0} & \small{50}  & \small{25} & \small{3.1}  \\
\small{18} & \small{0}  & \small{7.0}  & \small{150} & \small{25} & \small{4.1}  \\
\small{19} & \small{47} & \small{12.0} & \small{110} & \small{36} & \small{1.9}  \\
\small{20} & \small{0}  & \small{7.0}  & \small{150} & \small{45} & \small{8.3}  \\
\small{21} & \small{90} & \small{7.0}  & \small{50}  & \small{45} & \small{5.8}  \\
\small{22} & \small{63} & \small{7.0}  & \small{150} & \small{25} & \small{5.3}  \\
\small{23} & \small{42} & \small{12.0} & \small{50}  & \small{34} & \small{2.8}  \\
\small{24} & \small{36} & \small{12.0} & \small{110} & \small{25} & \small{4.0}  \\
\small{25} & \small{44} & \small{7.0}  & \small{100} & \small{45} & \small{5.5}  \\
\small{26} & \small{0}  & \small{7.0}  & \small{150} & \small{35} & \small{8.7}  \\
\small{27} & \small{32} & \small{7.0}  & \small{150} & \small{36} & \small{6.3}  \\
\small{28} & \small{0}  & \small{7.0}  & \small{90}  & \small{25} & \small{5.5}  \\
\bottomrule
\end{tabular}
\end{table}

The corresponding CMOD histories for all three replicate specimens are shown in Fig.~\ref{fig:representative_fracture_response}(b). In each test, the CMOD initially increases gradually and subsequently rises more rapidly as fracture is approached, followed by an abrupt, near-vertical increase upon unstable crack propagation. The final datum immediately preceding this abrupt increase, marked as the point of failure, defines the end of the deformation history used for evaluation of the J-integral. The complete loading and fracture process occurs within a fraction of a second, reflecting the brittle fracture response of the photopolymer in its glassy state. Despite some specimen-to-specimen variation in the precise failure time, the three CMOD histories exhibit similar overall evolution.

\begin{figure}[t!]
    \centering
    \includegraphics[width=4.475in]{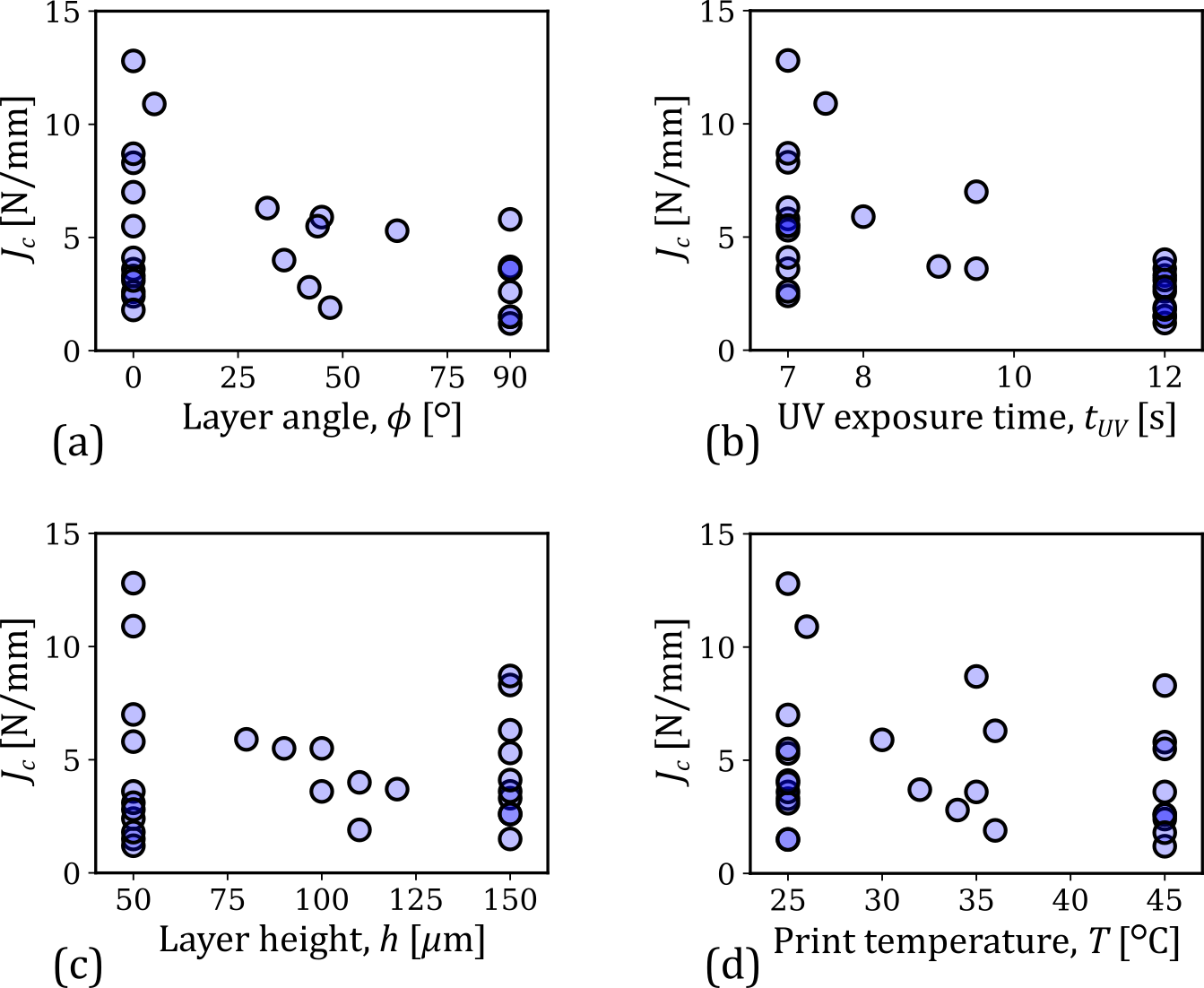}
    \caption{Measured fracture resistance as a function of individual processing parameters: (a) layer angle, $\phi$, (b) UV exposure time, $t_{UV}$, (c) layer height, $h$, and (d) print temperature, $T$. Each marker represents one experimentally evaluated processing condition, and the plotted response is the replicate-averaged critical J-integral, $J_c$. The broad distribution of points across the input ranges shows that the active learning campaign explored low, intermediate, and high values of each processing parameter.}
    \label{fig:jc_vs_inputs}
\end{figure}

Figure~\ref{fig:representative_fracture_response}(c) shows the corresponding J-integral histories computed from the measured load, CMOD, and DIC-derived hinge-point location. The individual responses exhibit a sigmoidal evolution, with a gradual initial increase in $J$, a more rapid intermediate rise, and an eventual approach toward a peak or near-asymptotic value immediately before crack propagation. For visualization of the average response and its variability, each individual J-integral history was extended as a constant beyond its respective failure point; these extensions are used only for constructing the average and standard-deviation envelope and do not represent post-fracture J-integral measurements. For each specimen, the critical J-integral was defined as the value at the final frame preceding visible crack propagation. The three replicate-specific critical values were then averaged, yielding $J_c=3.4$~N/mm for this processing condition. This replicate-averaged quantity is denoted by $J_c$ throughout the remainder of the manuscript and serves as the primary fracture resistance metric in the present work. Thus, each processing condition in the Bayesian active learning dataset is represented by a single experimentally derived $J_c$ value, linking the measured fracture resistance to the corresponding DLP processing parameters.

\subsection{Bayesian-active-learning-derived process--fracture mapping}
\label{subsection:process_fracture_mapping}

\paragraph{Exploration of the DLP processing space}

\begin{figure}[t!]
    \centering
    \includegraphics[width=5.190in]{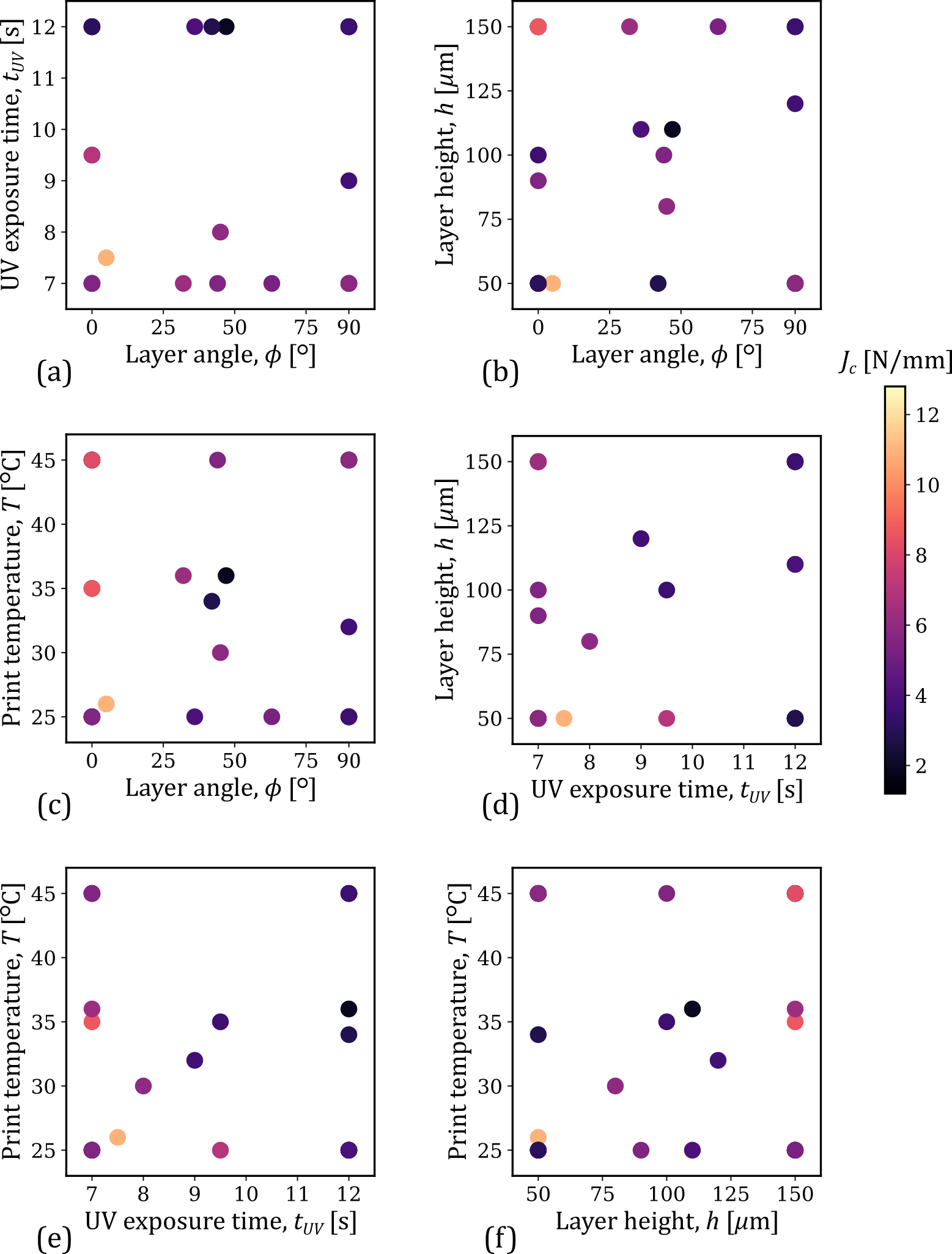}
    \caption{Pairwise projections of the explored four-dimensional DLP processing space: (a) $\phi$--$t_{UV}$, (b) $\phi$--$h$, (c) $\phi$--$T$, (d) $t_{UV}$--$h$, (e) $t_{UV}$--$T$, and (f) $h$--$T$. Each point represents one experimentally evaluated processing condition, and marker color denotes the measured replicate-averaged critical J-integral, $J_c$. The six projections illustrate the spread of the active-learning-selected experiments across the physical design space and highlight the multivariate character of the process--fracture relationship.}
    \label{fig:pairwise_exploration}
\end{figure}

The Bayesian-active-learning-guided experimental campaign evaluated 28 distinct DLP processing conditions, comprising two randomly selected initial conditions followed by 26 conditions selected sequentially through active learning (Section~\ref{sec:bayesian_active_learning}). For each processing condition, the replicate-averaged fracture resistance, $J_c$, was obtained using the DIC-assisted extraction procedure described in Section~\ref{subsection:DIC_J_c} and demonstrated for a representative experiment in the preceding subsection. The complete set of investigated processing conditions and corresponding measured $J_c$ values are listed in Table~\ref{tab:explored_conditions}. These data constitute the experimental basis for the process--fracture map developed in this work.

Figure~\ref{fig:jc_vs_inputs} shows the measured $J_c$ values as a function of each physical processing parameter, $\phi$, $t_{UV}$, $h$, and $T$. The selected experiments span low, intermediate, and high values of all four inputs, indicating that the active learning campaign explored a broad portion of the admissible design space rather than concentrating only in a narrow neighborhood of a single candidate optimum. This behavior is consistent with the exploration-biased nature of the modified UCB-style acquisition function adopted in this work. At the same time, the marginal plots show that the dependence of $J_c$ on any individual parameter is not fully described by a simple one-dimensional trend. Similar values of a given processing parameter can correspond to substantially different fracture resistances, indicating that the remaining processing variables and their coupled effects play an important role in determining $J_c$. Thus, the process--fracture relationship cannot be interpreted reliably from any individual processing parameter in isolation.

Figure~\ref{fig:pairwise_exploration} further visualizes the explored processing conditions through six pairwise projections of the four-dimensional design space. Each panel shows two physical processing parameters, while marker color denotes the measured replicate-averaged $J_c$. These projections confirm that the active learning campaign sampled near the boundaries as well as within the interior of the design space. The color variation within each projection further highlights the multivariate character of the process--fracture relationship: points that appear close in a two-parameter projection can exhibit different fracture resistances because they differ in the remaining two processing variables. Thus, Figs.~\ref{fig:jc_vs_inputs} and~\ref{fig:pairwise_exploration} together show that the proposed active learning campaign generated a sparse but broadly distributed experimental dataset for learning the process--fracture map.

\begin{figure}[t!]
    \centering
    \includegraphics[width=4.530in]{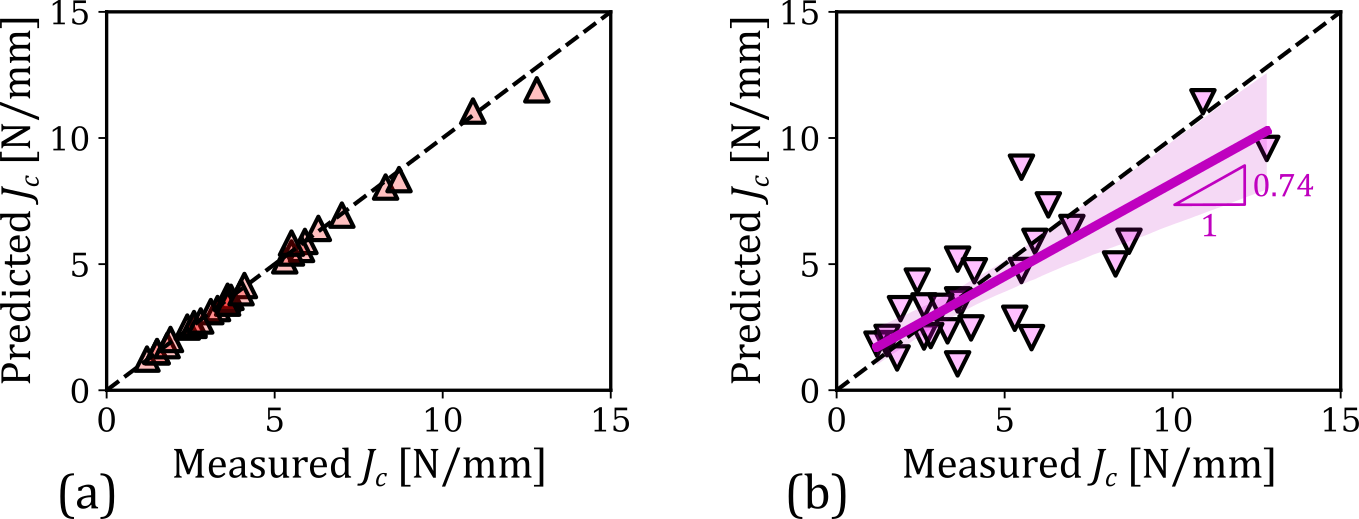}
    \caption{Performance of the final GPR process--fracture surrogate. (a) Training performance, showing the measured replicate-averaged critical J-integral, $J_c$, compared with the corresponding GPR-predicted values for all 28 experimentally evaluated processing conditions used to train the final surrogate. The dashed line denotes the ideal one-to-one prediction, and the high training coefficient of determination, $R^2=0.99$, indicates that the model closely fits the measured dataset. (b) Leave-one-out cross-validation (LOOCV) performance, where each point compares the measured $J_c$ for a withheld processing condition with the corresponding prediction from a GPR model trained on the remaining 27 conditions. The dashed line denotes the ideal one-to-one prediction, while the solid line denotes the linear regression fit to the LOOCV predictions; the shaded region denotes the corresponding 95\% confidence interval. The LOOCV results yield $R^2=0.63$, Pearson correlation coefficient $r=0.81$, and regression slope of 0.74, indicating that the model captures the dominant process--fracture trends while retaining finite prediction error in the sparse four-dimensional design space.}
    \label{fig:gpr_model_performance}
\end{figure}

\paragraph{\textbf{GPR surrogate performance and learned process--fracture map}}
After completion of the active learning campaign, the final GPR surrogate was trained using all 28 replicate-averaged experimental measurements. The nugget parameter was set to $\alpha=0.05$. This value was found to balance training accuracy and cross-validation performance, thereby reducing the likelihood of overfitting to the small experimental dataset while still allowing the surrogate to capture the observed multivariate process--fracture trends. The fitted model reproduced the measured training data with high accuracy, yielding a training coefficient of determination of $R^2=0.99$. This agreement is shown in Fig.~\ref{fig:gpr_model_performance}(a), where the GPR-predicted $J_c$ values for the experimentally evaluated processing conditions are compared with the corresponding measured values. The close clustering of the data around the one-to-one line indicates that the final surrogate provides a faithful interpolation of the available experimental dataset.

Because high training accuracy alone does not establish predictive reliability, the surrogate was further evaluated using leave-one-out cross-validation (LOOCV). In each LOOCV fold, one experimental datapoint was withheld, the GPR model was retrained using the remaining 27 datapoints, and the $J_c$ value for the withheld processing condition was predicted. The collection of left-out predictions across all 28 folds yielded an overall LOOCV coefficient of determination of $R^2=0.63$, a Pearson correlation coefficient of $r=0.81$, and a linear-regression slope of 0.74 between measured and predicted $J_c$ values. The corresponding parity plot is shown in Fig. \ref{fig:gpr_model_performance}(b). These results indicate that the surrogate captures the dominant trends in the process--fracture relationship, although with reduced predictive accuracy relative to the training fit. This reduction is expected given the limited number of experimental observations, the four-dimensional input space, experimental variability in fracture testing, and the complex, potentially nonlinear dependence of fracture resistance on the processing variables.

The LOOCV results also provide useful insight into the character of the learned model. The positive correlation and moderate $R^2$ value indicate that the GPR surrogate can distinguish lower- and higher-fracture-resistance regions of the explored process space, which is the primary requirement for subsequent sensitivity analysis and process interpretation. However, the regression slope below unity suggests that the cross-validated predictions somewhat compress the measured response range, underpredicting some high-$J_c$ cases and/or overpredicting some low-$J_c$ cases. Thus, the model should be interpreted as a data-efficient statistical map of the investigated process window rather than as a fully resolved deterministic description of the fracture response. Within this context, the surrogate provides a useful basis for identifying process trends and parameter importance from a modest experimental campaign.

The optimized Mat\'ern-$3/2$ kernel lengthscale (see Eq.~(\ref{eq:matern_kernel})) of the final GPR model was $\ell_{\mathrm{GP}}=0.58$ in normalized input space. Since the input variables were scaled to the unit hypercube, this lengthscale indicates that the learned response varies over an intermediate fraction of the design space: the map is neither nearly flat over the investigated domain nor so rapidly varying that only very local interpolation is possible. This intermediate smoothness is consistent with the trends observed in Figs.~\ref{fig:jc_vs_inputs}--\ref{fig:gpr_model_performance}, where the response exhibits clear process dependence but also substantial variability arising from the coupled dependence on multiple processing variables. Thus, the final trained GPR surrogate provides a continuous statistical representation of the process--fracture relationship within the investigated processing domain, which is interrogated in the following section through local one-at-a-time response analysis and global Sobol sensitivity analysis.

\subsection{Parameter importance, physical interpretation, and practical implications}
\label{subsection:sensitivity_results}

\paragraph{\textbf{One-at-a-time local sensitivity analysis}}
The final GPR process--fracture surrogate was analyzed using both one-at-a-time surrogate-based response curves and variance-based Sobol sensitivity indices. The one-at-a-time analysis provides a local, conditional view of how the predicted fracture resistance varies with each processing parameter near the design space center, whereas the Sobol analysis quantifies the relative global influence of each parameter over the full investigated process window. The resulting local response ranges (Eq. (\ref{eq:local_range})), local sensitivities (Eqs. (\ref{eq:local_sensitivity_normalized}--\ref{eq:local_sensitivity_physical})), and Sobol indices (Eqs. (\ref{eq:Sobol_first-order}--\ref{eq:Sobol_total-order})) are summarized in Table~\ref{tab:sensitivity_metrics}.

\begin{table}[b!]
\centering
\caption{Local and global sensitivity metrics obtained from the final GPR process--fracture surrogate. Here, $\Delta J_j$ is the one-at-a-time predicted response range, $m_j$ is the normalized local sensitivity at the design space center, $m_j^{\mathrm{phys}}$ is the corresponding physical-coordinate slope, and $S_j$ and $S_j^{\mathrm{tot}}$ are the first-order and total-order Sobol indices, respectively.}
\label{tab:sensitivity_metrics}
\begin{tabular}{lccccc}
\toprule
\small{\textbf{Processing parameter}} 
& \small{\textbf{$\Delta J_j$}} 
& \small{\textbf{$m_j$}} 
& \small{\textbf{$m_j^{\mathrm{phys}}$}} 
& \small{\textbf{$S_j$}} 
& \small{\textbf{$S_j^{\mathrm{tot}}$}} \\
& \small{\textbf{[N/mm]}} 
& \small{\textbf{[N/mm]}} 
& \small{\textbf{[physical units]}} 
& 
& \\
\midrule
\small{Layer angle, $\phi$} 
& \small{0.72} 
& \small{$-0.32$} 
& \small{$-0.0036$~(N/mm)/$^\circ$} 
& \small{0.0916} 
& \small{0.2208} \\

\small{UV exposure time, $t_{UV}$} 
& \small{3.69} 
& \small{$-4.64$} 
& \small{$-0.93$~(N/mm)/s} 
& \small{0.6780} 
& \small{0.7581} \\

\small{Layer height, $h$} 
& \small{0.26} 
& \small{$-0.065$} 
& \small{$-0.00065$~(N/mm)/$\mu$m} 
& \small{0.0015} 
& \small{0.1482} \\

\small{Print temperature, $T$} 
& \small{0.80} 
& \small{$-1.095$} 
& \small{$-0.055$~(N/mm)/$^\circ$C} 
& \small{0.0346} 
& \small{0.1877} \\
\bottomrule
\end{tabular}
\end{table}

\begin{figure}[t!]
    \centering
    \includegraphics[width=4.405in]{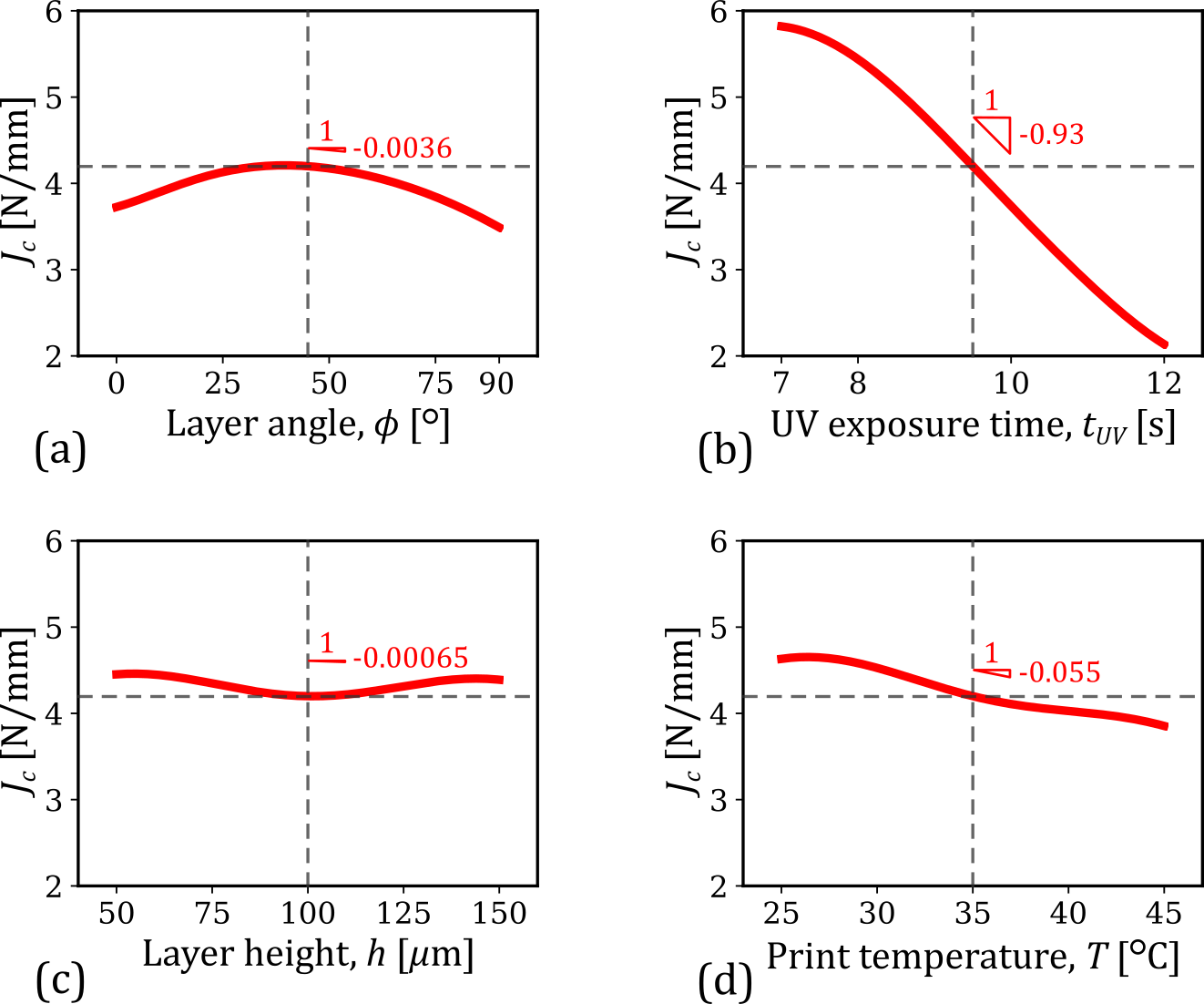}
    \caption{One-at-a-time surrogate-based response curves around the design space center. Each panel shows the GPR-predicted critical J-integral, $J_c$, as one processing parameter is varied across its admissible range while all remaining parameters are held fixed at the center of the normalized design space: $\tilde{\mathbf{X}}_c=[0.5,0.5,0.5,0.5]$. The response curves are shown for (a) layer angle, $\phi$, (b) UV exposure time, $t_{UV}$, (c) layer height, $h$, and (d) print temperature, $T$. Dashed horizontal and vertical lines denote the predicted response and parameter value at the design space center, respectively. The annotated slopes indicate the local physical-coordinate sensitivity $m_j^{\mathrm{phys}}$ at the design space center; the corresponding normalized sensitivities ($m_j$) used for cross-parameter comparison are reported in Table~\ref{tab:sensitivity_metrics}.}
    \label{fig:one_at_a_time_response}
\end{figure}

Figure~\ref{fig:one_at_a_time_response} shows the one-at-a-time response curves obtained by varying each processing parameter across its admissible range while holding the remaining parameters fixed at the design space center. At the design space center, the surrogate predicts a fracture resistance of $J_c=4.20$~N/mm, indicated by the common horizontal dashed line in all four panels. For direct visual comparison, the same vertical axis range, $J_c\in[2,6]$~N/mm, is used in all four panels. This range is smaller than the overall measured range of fracture resistance in the full experimental campaign, where the maximum measured $J_c$ exceeded 12~N/mm, because the one-at-a-time curves represent local conditional slices through the four-dimensional GPR surrogate rather than the full range of possible multivariable combinations. The conditional response curves also reveal a distinctly nonlinear process--fracture relationship, with pronounced curvature in the dependence on UV exposure time and print temperature, and nonmonotonic variations with layer angle and layer height.

To compare the local influence of different processing parameters, the local slopes were evaluated with respect to the normalized input coordinates, as described in Section~\ref{subsection:surrogate_sensitivity}. For clarity, we denote the sensitivities associated with the physical processing parameters as $m_{\phi}$, $m_{t_{UV}}$, $m_h$, and $m_T$, corresponding respectively to $m_1$, $m_2$, $m_3$, and $m_4$ in the notation of Section~\ref{subsection:surrogate_sensitivity}.
These normalized sensitivities have units of N/mm and represent the predicted change in $J_c$ associated with a unit change in the corresponding normalized processing parameter. Based on this metric, UV exposure time has the largest local influence on fracture resistance, with $m_{t_{UV}}=-4.64$~N/mm. As shown in Fig.~\ref{fig:one_at_a_time_response}(b), $J_c$ decreases monotonically with increasing $t_{UV}$ when all other parameters are fixed at the design space center, and the predicted $J_c$ variation across the investigated exposure-time range is 3.69~N/mm. This strong negative trend suggests that, within the investigated process window, longer UV exposure tends to reduce fracture resistance. A plausible interpretation is that increased exposure promotes a higher degree of cure and crosslinking, which may increase brittleness and reduce the material's resistance to crack initiation and propagation under the present glassy, room-temperature testing conditions.

Print temperature has the second-largest local sensitivity, with $m_T=-1.095$~N/mm. As shown in Fig.~\ref{fig:one_at_a_time_response}(d), increasing $T$ from 25 to 45~$^\circ$C decreases the predicted $J_c$ by approximately 0.80~N/mm. This trend indicates that the local thermal environment during printing affects the fracture response, potentially through its influence on resin viscosity, recoating behavior, cure uniformity, and interlayer bonding. However, compared with UV exposure time, print temperature acts as a secondary factor in the local response around the design space center.

Layer angle and layer height exhibit weaker one-at-a-time effects near the design space center. For layer angle, the normalized local sensitivity is $m_{\phi}=-0.32$~N/mm, and the predicted response varies by 0.72~N/mm across the full range from $0^\circ$ to $90^\circ$. The response is nonmonotonic and approximately symmetric about the design space center, suggesting that the influence of layer angle is not captured by a simple increasing or decreasing trend in this local slice. Layer height produces the smallest local sensitivity, $m_h=-0.065$~N/mm, and the smallest local response range, 0.26~N/mm. The weak and nonmonotonic local dependence of $J_c$ on $\phi$ and $h$ indicates that these parameters have limited isolated influence near the design space center, although they may still contribute through interactions with other processing variables.

\begin{figure}[t!]
    \centering
    \includegraphics[width=4.892in]{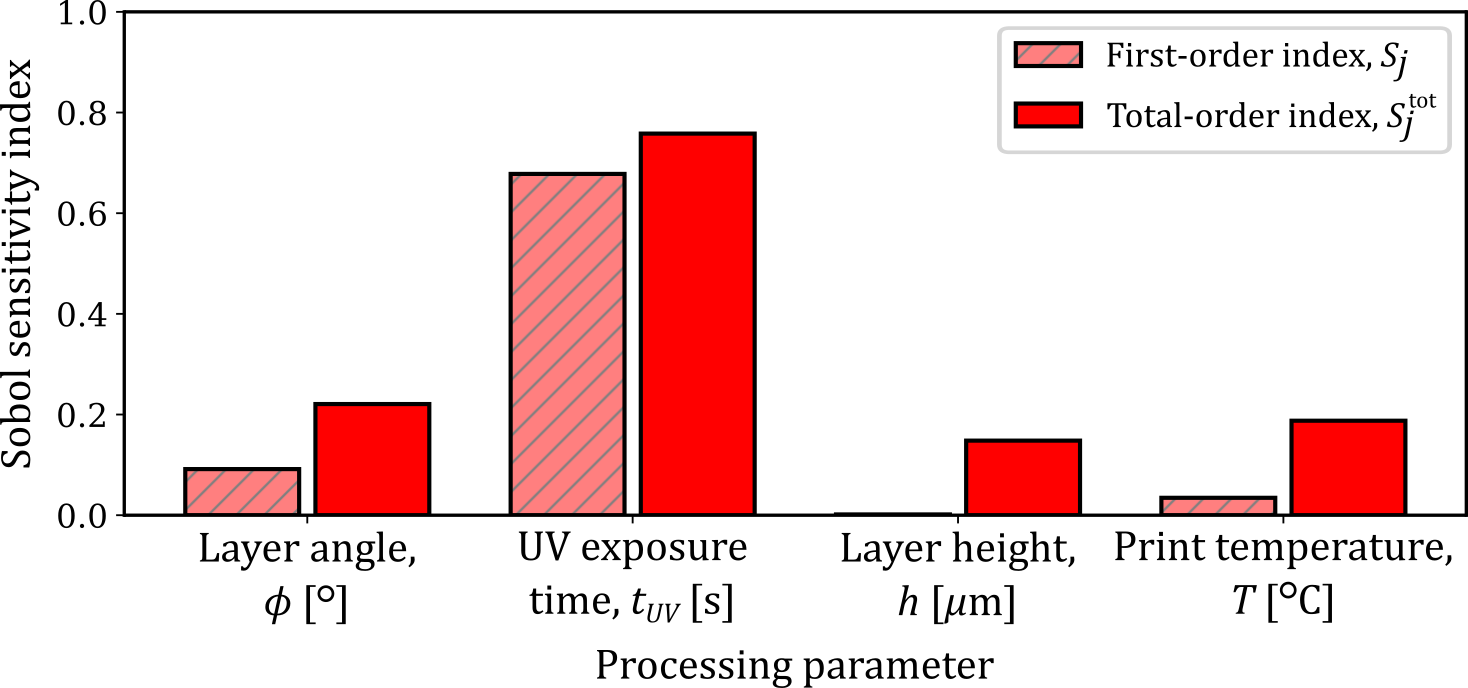}
    \caption{Variance-based global sensitivity analysis of the GPR process--fracture surrogate. The grouped bar graph compares the first-order Sobol index, $S_j$, and total-order Sobol index, $S_j^{\mathrm{tot}}$, for each DLP processing parameter: layer angle $\phi$, UV exposure time $t_{UV}$, layer height $h$, and print temperature $T$. The first-order index quantifies the direct contribution of a parameter to the variance in the surrogate-predicted critical J-integral, $J_c$, whereas the total-order index quantifies its overall contribution, including both direct effects and interactions with other processing parameters. Differences between $S_j$ and $S_j^{\mathrm{tot}}$ indicate the extent to which a parameter participates in interaction effects.}
    \label{fig:sobol_indices}
\end{figure}

\paragraph{\textbf{Global Sobol sensitivity analysis}}
The global Sobol sensitivity analysis, shown in Fig.~\ref{fig:sobol_indices}, provides a complementary view of parameter importance over the full investigated process window. The first-order Sobol indices are $S_{\phi}=0.0916$, $S_{t_{UV}}=0.6780$, $S_h=0.0015$, and $S_T=0.0346$. Thus, UV exposure time is by far the dominant direct contributor to the variance in the surrogate-predicted $J_c$, accounting for approximately 67.8\% of the output variance by its direct effect alone. The next largest direct contribution comes from layer angle, followed by print temperature, while layer height has an almost negligible first-order effect. The sum of the first-order indices is 0.8058, indicating that approximately 80.6\% of the response variance is explained by direct additive effects of the four parameters, while the remaining approximately 19.4\% is associated with interaction effects and higher-order contributions.

The total-order Sobol indices further clarify the role of interactions. The total-order indices are $S_{\phi}^{\mathrm{tot}}=0.2208$, $S_{t_{UV}}^{\mathrm{tot}}=0.7581$, $S_h^{\mathrm{tot}}=0.1482$, and $S_T^{\mathrm{tot}}=0.1877$. Based on total-order influence, the parameters rank as
\[
t_{UV} > \phi > T > h.
\]
This ranking confirms that UV exposure time remains the most influential parameter even after interactions are included. However, the difference between total-order and first-order indices is nonzero for all four parameters, indicating that each parameter participates in interaction effects to some extent. The interaction-associated differences, $S_j^{\mathrm{tot}}-S_j$, are largest for print temperature, layer height, and layer angle, and smallest for UV exposure time. This suggests that the influence of UV exposure time is primarily direct, whereas the effects of print temperature, layer height, and layer angle are more strongly expressed through coupling with other processing variables.

Together, the local and global sensitivity analyses provide a consistent interpretation of the learned process--fracture map. UV exposure time is the primary processing parameter governing fracture resistance in the investigated DLP photopolymer system, with lower exposure times generally associated with higher predicted $J_c$ values within the investigated processing domain. Print temperature has a weaker but still meaningful influence, particularly through both local monotonic trends and interaction effects. Layer angle and layer height have relatively small isolated effects near the design space center, but their non-negligible total-order indices indicate that they should not be ignored in multivariable process design. These results support the central premise of the present study: sparse, Bayesian-active-learning-guided experiments combined with surrogate-based sensitivity analysis can reveal both dominant processing variables and interaction-driven effects that would otherwise require substantially larger full-factorial experimental campaigns and could be obscured when only a limited subset of input variables is considered.
 
\section{Summary and Conclusions}
\label{sec:summary_and_conclusion}

This study investigated the use of Bayesian active learning for data-efficient process--fracture mapping of a DLP-printed photopolymer. The four-dimensional processing space considered layer angle, UV exposure time, layer height, and print temperature, for which exhaustive experimentation would require more than 230,000 distinct parameter combinations at the prescribed processing resolutions. To explore this space efficiently, a GPR surrogate was coupled with a modified UCB-style acquisition function to sequentially select new experiments. Fracture resistance was quantified using the replicate-averaged critical J-integral, $J_c$, obtained from three-point-bending experiments and DIC-assisted evaluation of the specimen deformation and hinge-point kinematics. Starting from two randomly selected conditions, the active learning campaign evaluated a total of only 28 distinct processing conditions, each using three replicate specimens.

The resulting GPR surrogate provided a continuous statistical representation of the process--fracture relationship and reproduced the training data with $R^2=0.99$. Leave-one-out cross-validation yielded $R^2=0.63$ and a Pearson correlation coefficient of $r=0.81$, indicating that the surrogate captured the dominant variation in fracture resistance despite the sparsity of the four-dimensional experimental dataset. Surrogate-based sensitivity analysis further revealed a distinctly nonlinear and multivariable process--fracture relationship. UV exposure time was the dominant processing parameter, with first-order and total-order Sobol indices of $S_{t_{UV}}=0.6780$ and $S_{t_{UV}}^{\mathrm{tot}}=0.7581$, respectively, and lower exposure times were generally associated with higher predicted fracture resistance within the investigated processing domain. Based on total-order influence, the parameters ranked as $t_{UV}>\phi>T>h$. The first-order indices summed to 0.8058, indicating that approximately 19.4\% of the predicted response variance was associated with interactions and higher-order effects; these interaction contributions were particularly important for layer angle, layer height, and print temperature.

Overall, the results demonstrate that Bayesian-active-learning-guided experimentation combined with GPR-based process mapping and surrogate sensitivity analysis can extract useful process--property relationships from a comparatively small experimental campaign. In addition to identifying the dominant processing variable, the framework revealed nonlinear conditional trends and interaction-driven effects that would be difficult to resolve by considering only a limited subset of processing variables and would otherwise require substantially larger full-factorial experimental campaigns. The present study therefore provides a general data-efficient framework for experimentally exploring high-dimensional processing spaces and identifying the processing parameters and interactions governing fracture resistance in additively manufactured polymers.
 
\section*{Acknowledgments} \label{Acknowledgments}

This work was supported in part by the Louisiana Experimental Program to Stimulate Competitive Research (EPSCoR) through the NSF EPSCoR LAMDA Seed Funding Track 1B initiative (Contract No. NSF(2023)-LAMDATr1B-17), funded by the National Science Foundation under Cooperative Agreement No. OIA-1946231 and the Board of Regents Support Fund; and additionally by the NASA EPSCoR Rapid Response Research (R3) Program (Contract No. NASA(2024-25)-Rapid-12), under NASA Cooperative Agreement No. 80NSSC24M0126 and the Board of Regents Support Fund.




\appendix
\setcounter{table}{0}
\renewcommand{\thetable}{A.\arabic{table}}
\renewcommand{\theequation}{A.\arabic{equation}}
\setcounter{equation}{0}

\section{Numerical estimation of Sobol sensitivity indices}
\label{app:sobol}

This appendix summarizes the numerical procedure used to estimate first-order and total-order Sobol sensitivity indices from the final GPR surrogate model. The purpose of this analysis is to quantify the relative influence of each DLP processing parameter on the replicate-averaged critical J-integral, $J_c$.

\paragraph{Sampling strategy}
Let the surrogate-predicted response be denoted by
\begin{equation}
Y=\mathcal{M}(\tilde{\mathbf{X}}),
\end{equation}
where $\tilde{\mathbf{X}}\in[0,1]^4$ is the normalized vector of input processing parameters and $\mathcal{M}$ is the final trained GPR model. The normalized variables are obtained from the physical processing parameters using the min--max transformation defined in Eq.~(\ref{eq:minmax_scaling}). Because this transformation is linear and one-to-one, sensitivity indices computed with respect to the normalized variables correspond directly to the physical processing parameters $\phi$, $t_{UV}$, $h$, and $T$.

To estimate Sobol indices, two independent normalized input sample matrices,
\begin{equation}
\tilde{\mathbf{A}},\tilde{\mathbf{B}}\in[0,1]^{N\times d},
\end{equation}
were generated, where $d=4$ is the number of input parameters and $N=50000$ is the Monte Carlo sample size. Each row of $\tilde{\mathbf{A}}$ and $\tilde{\mathbf{B}}$ corresponds to one sampled normalized processing-condition vector. The samples were generated using Latin hypercube sampling \cite{Iman2008}, assuming independent uniform distributions for all four normalized inputs. Because the min--max transformation in Eq.~(\ref{eq:minmax_scaling}) is a linear one-to-one transformation of the physical parameter ranges listed in Table~\ref{tab:processing_parameters}, this sampling is equivalent to uniform sampling over the investigated physical process window followed by normalization.

For each parameter $j\in\{1,\dots,d\}$, a mixed matrix $\tilde{\mathbf{A}}_{\tilde{\mathbf{B}}}^{(j)}$ was constructed by replacing the $j$-th column of $\tilde{\mathbf{A}}$ with the corresponding column of $\tilde{\mathbf{B}}$ while keeping all remaining columns from $\tilde{\mathbf{A}}$. In component form,
\begin{equation}
\left(\tilde{\mathbf{A}}_{\tilde{\mathbf{B}}}^{(j)}\right)_{ik}
=
\begin{cases}
\tilde{B}_{ik}, & k=j,\\
\tilde{A}_{ik}, & k\neq j,
\end{cases}
\qquad i=1,\dots,N,\;\;k=1,\dots,d.
\end{equation}

\paragraph{Surrogate-model evaluations}
The trained GPR surrogate was evaluated on the rows of $\tilde{\mathbf{A}}$, $\tilde{\mathbf{B}}$, and $\tilde{\mathbf{A}}_{\tilde{\mathbf{B}}}^{(j)}$ to obtain
\begin{equation}
y_{\tilde{A}}^{(i)}
=
\mathcal{M}\!\left(\tilde{\mathbf{A}}^{(i)}\right),
\qquad
y_{\tilde{B}}^{(i)}
=
\mathcal{M}\!\left(\tilde{\mathbf{B}}^{(i)}\right),
\qquad
y_{\tilde{A}_{\tilde{B}}^{(j)}}^{(i)}
=
\mathcal{M}\!\left(
\left(\tilde{\mathbf{A}}_{\tilde{\mathbf{B}}}^{(j)}\right)^{(i)}
\right),
\qquad i=1,\dots,N,
\end{equation}
where the superscript $(i)$ denotes the $i$-th row of the corresponding matrix.

For convenience, an empirical output-variance estimator was defined as
\begin{equation}
\label{eq:Vhat_appendix}
\widehat{V}_j
=
\frac{1}{2N}\sum_{i=1}^{N}
\left[
\left(y_{\tilde{B}}^{(i)}\right)^2
+
\left(y_{\tilde{A}_{\tilde{B}}^{(j)}}^{(i)}\right)^2
\right]
-
\left[
\frac{1}{2N}\sum_{i=1}^{N}
\left(
y_{\tilde{B}}^{(i)}
+
y_{\tilde{A}_{\tilde{B}}^{(j)}}^{(i)}
\right)
\right]^2.
\end{equation}
This quantity provides a consistent estimate of the common output variance $\mathrm{Var}(Y)$ and was used to normalize the first-order and total-order indices for parameter $j$.

\paragraph{First-order Sobol index}
The first-order Sobol index quantifies the fraction of output variance that can be attributed to a single input parameter acting alone, independent of interaction effects. Using the pick-freeze formulation of Monod et al.~\cite{Monod2006}, the first-order sensitivity index for parameter $j$ was estimated as
\begin{equation}
\label{eq:Sj_appendix}
S_j
=
\frac{
\frac{1}{N}\sum_{i=1}^{N}
y_{\tilde{B}}^{(i)}\,
y_{\tilde{A}_{\tilde{B}}^{(j)}}^{(i)}
-
\left[
\frac{1}{2N}\sum_{i=1}^{N}
\left(
y_{\tilde{B}}^{(i)}
+
y_{\tilde{A}_{\tilde{B}}^{(j)}}^{(i)}
\right)
\right]^2
}{
\widehat{V}_j
}.
\end{equation}
A larger value of $S_j$ indicates that the output variance is more strongly influenced by the direct effect of the corresponding input parameter.

\paragraph{Total-order Sobol index}
The total-order Sobol index measures the overall importance of a parameter, including both its direct contribution and all interaction effects involving that parameter. In the present study, total-order indices were computed using Jansen's estimator \cite{Jansen1999,Ghanem2017},
\begin{equation}
\label{eq:STj_appendix}
S^{\mathrm{tot}}_j
=
\frac{
\frac{1}{2N}\sum_{i=1}^{N}
\left(
y_{\tilde{A}}^{(i)}
-
y_{\tilde{A}_{\tilde{B}}^{(j)}}^{(i)}
\right)^2
}{
\widehat{V}_j
}.
\end{equation}
The paired input vectors $\tilde{\mathbf{A}}^{(i)}$ and $\left(\tilde{\mathbf{A}}_{\tilde{\mathbf{B}}}^{(j)}\right)^{(i)}$ share all input coordinates except the $j$-th coordinate. Their output difference therefore captures the variation associated with changing parameter $j$, including its interactions with the remaining inputs. When $S^{\mathrm{tot}}_j$ substantially exceeds $S_j$, the difference reflects the presence of non-negligible interaction effects involving parameter $j$.

\paragraph{Interpretation}
For the present process--fracture problem, the first-order index $S_j$ indicates how much of the variability in the surrogate-predicted critical J-integral can be explained by varying a single processing parameter alone. In contrast, the total-order index $S^{\mathrm{tot}}_j$ reflects the full importance of that parameter, including its coupling with the remaining processing variables. Accordingly, comparison of first-order and total-order indices provides a compact quantitative picture of both the direct and interaction-driven contributions of layer angle, UV exposure time, layer height, and print temperature to the fracture resistance of the DLP-printed photopolymer.

 \bibliographystyle{elsarticle-num} 
 \bibliography{cas-refs}





\end{document}